\documentclass[prd,aps,nofootinbib,floatfix,10pt,preprint]{revtex4}
\usepackage{amsmath,graphicx,epsfig,amssymb,dsfont,mathtools}
\usepackage[usenames]{color}
\usepackage{mathrsfs}
\usepackage{ulem} 
\usepackage{bigstrut}
\usepackage{slashed}
\usepackage{multirow}
\usepackage{subfigure}

\begin{document}
\preprint{ACFI-T20-09, DESY 20-129}
\title{Transverse Energy-Energy Correlations of jets in the electron-proton Deep Inelastic Scattering at  HERA}

\author{Ahmed Ali\footnote{Email:ahmed.ali@desy.de}}
\affiliation{
Deutsches Elektronen-Synchrotron DESY, D-22607 Hamburg, Germany
}
\author{Gang Li\footnote{Corresponding Author. Email:ligang@umass.edu}}
\affiliation{
Amherst Center for Fundamental Interactions, Department of Physics, University of Massachusets Amherst,
MA 01003, USA
}
\author{Wei Wang\footnote{Email:wei.wang@sjtu.edu.cn}}
\affiliation{INPAC, Shanghai Key Laboratory for Particle Physics and Cosmology, MOE Key Laboratory for Particle Astrophysics and Cosmology, School of Physics and Astronomy, Shanghai Jiao Tong University, Shanghai 200240, China
}
\author{Zhi-Peng Xing\footnote{Corresponding Author. Email: zpxing@sjtu.edu.cn}}
\affiliation{INPAC, Shanghai Key Laboratory for Particle Physics and Cosmology, MOE Key Laboratory for Particle Astrophysics and Cosmology, School of Physics and Astronomy, Shanghai Jiao Tong University, Shanghai 200240, China
}

\begin{abstract}
We study the event shape variables,  transverse energy-energy correlation TEEC $(\cos \phi)$ and its asymmetry ATEEC $(\cos \phi)$ in
deep inelastic scattering (DIS) at the electron-proton collider HERA, where $\phi$ is the angle between two jets defined using a transverse-momentum $(k_T)$ jet algorithm.  At HERA, jets are defined in the Breit frame, and the leading nontrivial  transverse energy-energy  correlations  arise from the 3-jet configurations.  With the help of the   NLOJET++,  these functions are calculated  in the leading order (LO) and the next-to-leading order (NLO) approximations in QCD at the electron-proton center-of-mass energy $\sqrt{s}=314$ GeV.   We restrict the angular region to $-0.8 \leq \cos \phi \leq 0.8$, as the forward- and backward-angular regions require resummed logarithmic corrections, which we have neglected in this work. Following experimental jet-analysis at HERA,
 we restrict the DIS-variables $x$, $y=Q^2/(x s)$, where $Q^2=-q^2$ is the negative of the momentum transfer squared $q^2$, to $0  \leq x \leq 1$, $0.2 \leq y \leq 0.6$, and  the pseudo-rapidity variable in the
laboratory frame $(\eta^{\rm {lab}})$ to the range $-1 \leq \eta^{\rm {lab}} \leq 2.5$. The TEEC and ATEEC
functions are worked out for two ranges in $Q^2$, defined by $5.5~{\rm GeV}^2 \leq Q^2 \leq 80~{\rm GeV}^2$, called the low-$Q^2$-range, and $150~{\rm GeV}^2 \leq Q^2 \leq 1000~{\rm GeV}^2$, called the high-$Q^2$-range.
 We show the sensitivity of these functions on the parton distribution  functions (PDFs), the factorization $(\mu_F)$ and renormalization $(\mu_R)$ scales, and  on $\alpha_s(M_Z^2)$. Of these the correlations are stable against varying the scale $\mu_F$ and the PDFs, but they do depend on $\mu_R$. For the choice   of the scale $\mu_R= \sqrt{\langle E_T\rangle^2 +Q^2}$,
 advocated in earlier jet analysis at HERA, the shape variables TEEC and ATEEC are
 found perturbatively robust. These studies are useful  in the analysis of the HERA data, including the determination of $\alpha_s(M_Z^2)$ from the shape variables.
\end{abstract}
\maketitle

\section{Introduction}

Event shape variables involving the energy-momentum variables of hadrons and jets have played a crucial role in testing Quantum Chromodyamics (QCD), providing a detailed comparison with the experimentally measured shapes in high energy collisions and in determining the strong interaction coupling constant $\alpha_s(Q^2)$. Of these, the energy-energy correlation (EEC) and its asymmetry  (AEEC), introduced by Basham {\it et al.}  in $e^+e^-$ annihilation~\cite{Basham:1978bw,Basham:1978zq} have received  a lot of experimental and theoretical attention. Next-to-leading order (NLO) corrections in   $\alpha_s(Q^2)$ were calculated long ago for the EEC in $e^+e^-$
annihilation, using a number of different methods to regulate the soft and collinear divergences~\cite{Ali:1982ub,Barreiro:1986si,Richards:1982te,Richards:1983sr,Glover:1994vz,Schneider:1983iu,Falck:1988gb,Kramer:1996qr}. Accurate numerical results for the EEC are available from the program Event 2, based on the dipole subtraction technique~\cite{Catani:1996jh,Catani:1996vz}.
EEC has also been calculated to NNLO accuracy in perturbative
 QCD~\cite{DelDuca:2016csb,Tulipant:2017ybb} and in the next-to-next-to-leading logarithms (NNLL)~\cite{deFlorian:2004mp}.
 Recent advances in theoretical calculational techniques have led to a  renaissance of interest in this topic. In particular, an analytic NLO calculation of the EEC  in $e^+e^-$ annihilation~\cite{Dixon:2018qgp,Luo:2019nig}, and an all-order factorization formula for the EEC in the back-to-back limit~\cite{Moult:2018jzp,Dixon:2019uzg,Kologlu:2019mfz,Korchemsky:2019nzm},  are now available.
 We also mention here the derivation of the EEC function in the maximally supersymmetric $N=4$ super-Yang-Mills theory
 in the NLO accuracy~\cite{Belitsky:2013ofa}, which has been recently extended up to NNLO
 accuracy~\cite{Henn:2019gkr}. Experimental measurements of EEC in $e^+e^-$ annihilation are discussed
 in~\cite{Berger:1980yh,Behrend:1982na,Adeva:1991vw,Abreu:1990us,Abe:1994wv}.

  Following EEC in $e^+ e^-$ annihilation,
 transverse energy-energy correlation (TEEC) and the corresponding asymmetry (ATEEC)  were introduced in hadronic collisions
 at the SP$\bar {\rm P}$S~\cite{Ali:1984yp}, but did not evoke much experimental interest.  With the advent of the LHC era,  NLO corrections were calculated in $pp$ collisions~\cite{Ali:2012rn}.  They have been used by the ATLAS collaboration for comparison with data and in the determination of $\alpha_s(M_Z^2)$ from these shape functions~\cite{ATLAS:2015yaa,Aaboud:2017fml}. Recently,  TEEC in the dijet back-to-back limit in hadronic collisions has been derived, achieving an impressive  perturbative simplicity~\cite{Gao:2019ojf}.
Currently the TEEC-data  in $pp$ collisions are restricted in their theoretical interpretation  to NLO accuracy.

What concerns deep inelastic
scattering (DIS), event shape variables have also received a lot of theoretical attention ~\cite{Dasgupta:2003iq,Antonelli:1999kx,Dasgupta:2002dc,Kang:2013lga,Kang:2015swk,Gehrmann:2019hwf}.
Prominent among them are the thrust-distribution, 1-jettiness, jet-broadening, and the $C$ parameter, which have been calculated to
very high accuracy in fixed order (NNLO)~\cite{Gehrmann:2019hwf},  and in the resummed leading logarithms ($N^3LL$)~\cite{Kang:2015swk}. Some of these event shape variables have been measured by the
H1~\cite{Aktas:2005tz} and ZEUS~\cite{Chekanov:2002xk} collaborations at HERA. The definitions of these shape variables together with some others, such as the jet shape, can be seen in~\cite{Newman:2013ada}, where DIS and photoproduction experiments at HERA are reviewed.  However, to the best of our knowledge,
the transverse energy-energy correlation between the final state jets in deep inelastic scattering  has neither  been calculated nor
measured so far. Analogous to the TEEC for hadronic collisions~\cite{Ali:1984yp,Ali:2012rn}, TEEC in DIS  is introduced  in Eq.~(\ref{teec}) in the next section. It involves transverse energy correlations in two jets, defined by a jet-definition and jet algorithm, separated by an azimuthal angle $\phi$. We calculate TEEC and its asymmetry in DIS at HERA under realistic experimental conditions.

Jets  at HERA  are defined in the  Breit frame, in which  the exchanged  photon is at rest and the incoming and outgoing quarks are along the $z$ direction.  In this frame,  the involved  hadronic final states have zero  total transverse  momentum, and thus the leading nontrival transverse energy-energy correlation  comes from the 3-jet  configurations.   To match the measurements of jets at HERA, we  adopt the transverse-momentum $(k_T)$  algorithm to classify the jets~\cite{Catani:1993hr} and calculate the
 TEEC and its asymmetry (ATEEC) in the kinematic conditions
employed typically in H1 and ZEUS. The calculations are done in the NLO accuracy in the central angular
region, $-0.8 \leq \cos \phi \leq 0.8$.  This avoids the back-to-back angular configuration, i.e., near $\phi = \pi$, where  the leading logs (LL) and the
 next-to-leading logs (NLL), $\alpha_s^m(\mu) \ln ^n \tau$ ($ m \leq n)$  in the variable $\tau=\ln (1 + \cos \phi)/2$, have to be resummed. For the fixed-order perturbative calculations, we have
 used  the NLOJET++ package~\cite{Nagy:2001fj,Nagy:2001xb} and have tested it against the distributions obtained by Madgraph~\cite{Alwall:2014hca}. To achieve numerical stability, we have generated $10^9$ DIS events at HERA ($\sqrt{s}=314$ GeV), allowing us to reach an statistical accuracy of a few  percent  over  most of the phase space.

 Being weighted by the product of transverse energies
 of jets, both the TEEC and ATEEC are expected to be insensitive to the parton distribution functions (PDFs). To quantify this,
  we use two PDF sets of relatively recent vintage, the
  CT18~\cite{Hou:2019efy}, and MMHT14~\cite{Harland-Lang:2014zoa}. The main theoretical uncertainty in the jet physics comes
  from the scale-dependence, of these the so-called  factorization scale $\mu_F$ enters through the PDFs, and the partonic matrix elements depend essentially  on the renormalization scale $\mu_R$.  Detailed  studies  done for the inclusive jet and dijet data at HERA show that
  the $\mu_F$-dependence of the cross sections is small, but the $\mu_R$-dependence is substantial in the NLO accuracy~\cite{Andreev:2016tgi,Andreev:2017vxu}. We study these dependencies in TEEC and ATEEC, following the choice of the nominal scale, $\mu_0= \sqrt{\langle E_T \rangle^2 + Q^2}$,  where $\langle E_T\rangle$ denotes the average of $E_T$, as  advocated in these papers. Varying the
 scales in the range $\mu_{F}= (0.5, 2) \mu_0$,  we find that the $\mu_F$-dependence is small in the TEEC, not exceeding $5\%$
 over the $\cos \phi$ range, but the $\mu_R$-dependence is found to be significant.
  Thus, NNLO improvements are needed to reduce the $\mu_R$-uncertainty. However, fitting the HERA data on TEEC may also effectively reduce the allowed $\mu_R$-range.
  Finally, we show the sensitivity of the TEEC and ATEEC on the strong coupling constant $\alpha_s(M_Z^2)$,
for three representative values $\alpha_s(M_Z)= 0.108, 0.118, 0.128$. With the nominal choice of the scales
$\mu_F=\mu_R=\mu_0$, and the current central value of $\alpha_s(M_Z)=0.118$~\cite{Zyla:2020zbs}, we show that the
differential distributions TEEC$(\cos \phi)$
and ATEEC$(\cos \phi)$ are remarakbly stable perturbatively in both the $Q^2$-ranges. This remains to be tested in the NNLO accuracy.
 Our study presented here makes a good case for using the TEEC  in DIS-data as a precision test of perturbative QCD, following similar anayses done for the high energy $pp$ data at the LHC.

The rest of this paper is organized as follows. Section II collects the
definitions  of TEEC and its asymmetry. Experimental cuts to calculate these functions are stated in this section together with the jet algorithm used and the jet definitions.
 In Sec. III, we present the numerical  results calculated at next-to-leading order in $\alpha_s$  and estimate the uncertainty in the shape variables TEEC and ATEEC arising from the different PDFs, and the scale-dependence by varying the scale $\mu_F$ and $\mu_R$.  Of these, the $\mu_R$-dependence is substantial. Fixing the scale $\mu_R$ to the nominal value $\mu_0~,$  which provides a good fit of the inclusive-jet and dijet data at HERA~\cite{Andreev:2016tgi,Andreev:2017vxu}, we show the sensitvity of
 the TEEC and ATEEC on $\alpha_s(M_Z^2)$. A comparison of the LO and NLO results is also presented here.
 We summarise our results in the last section. A check of the NLOJET++ calculation is shown in Appendix-A at the LO, by using the package
 \text{MadGraph5\_aMC@NLO}~\cite{Alwall:2014hca}  with the MMHT14  PDF set.

\section{transverse energy-energy correlation and its asymmetry}\label{sec2}
\label{sec:EEC definitions}

\begin{figure}[htbp!]
   \begin{minipage}[t]{0.3\linewidth}
  \centering
  \includegraphics[width=0.8\columnwidth]{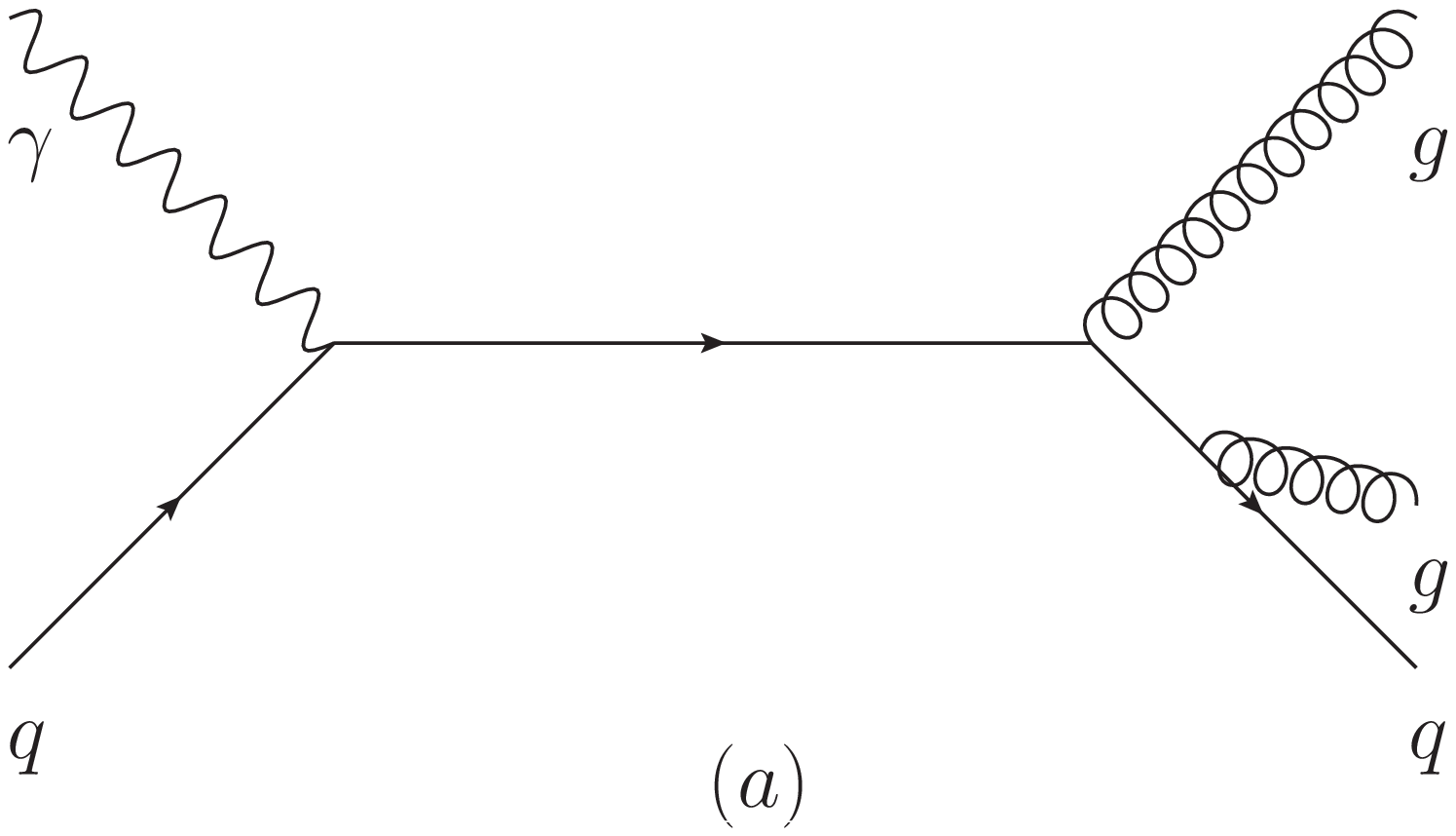}
  \end{minipage}
   \begin{minipage}[t]{0.3\linewidth}
  \centering
  \includegraphics[width=0.8\columnwidth]{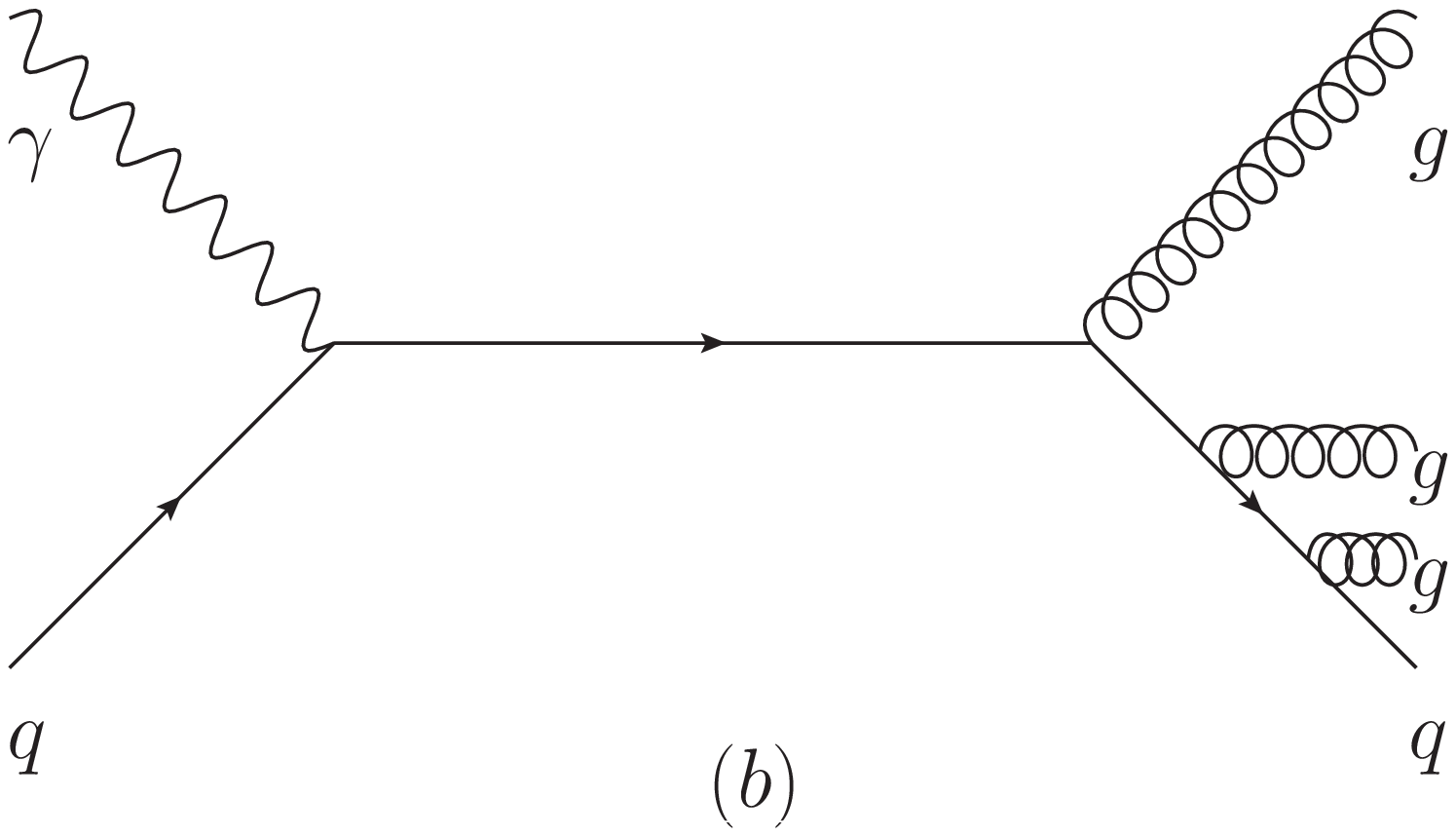}
  \end{minipage}
  \begin{minipage}[t]{0.3\linewidth}
  \centering
  \includegraphics[width=0.8\columnwidth]{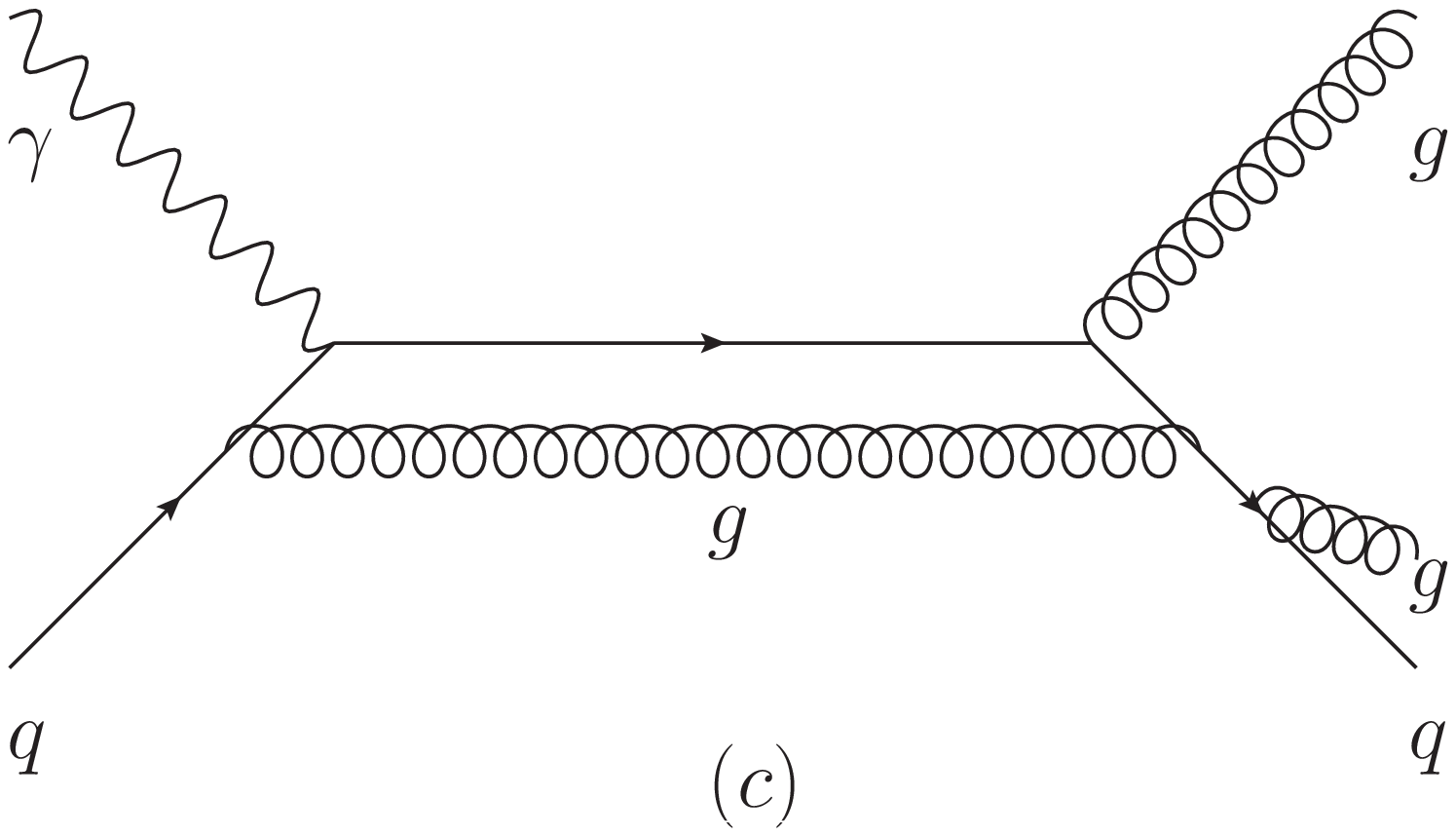}
  \end{minipage}
  \begin{minipage}[t]{0.3\linewidth}
  \centering
  \includegraphics[width=0.8\columnwidth]{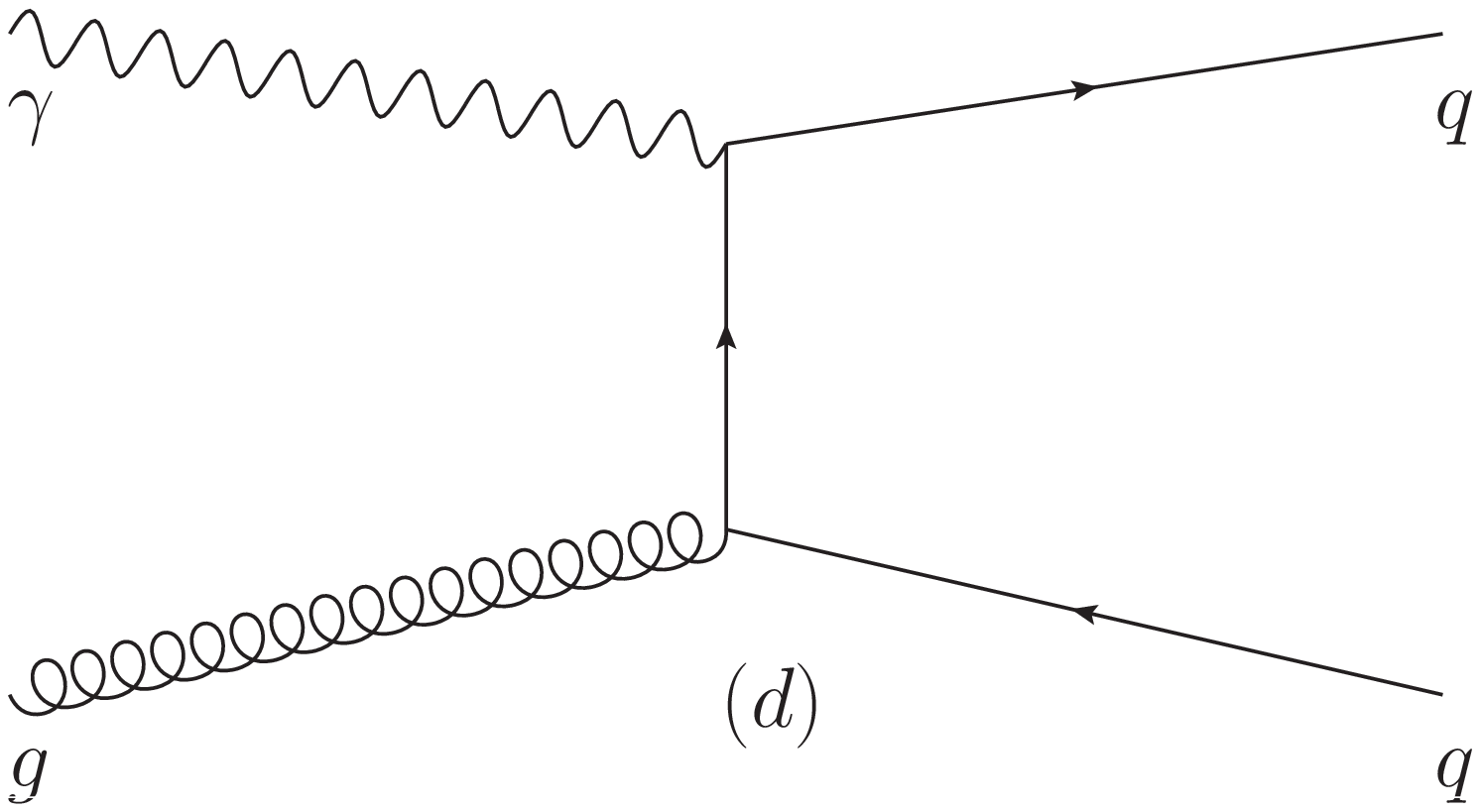}
  \end{minipage}
   \begin{minipage}[t]{0.3\linewidth}
  \centering
  \includegraphics[width=0.8\columnwidth]{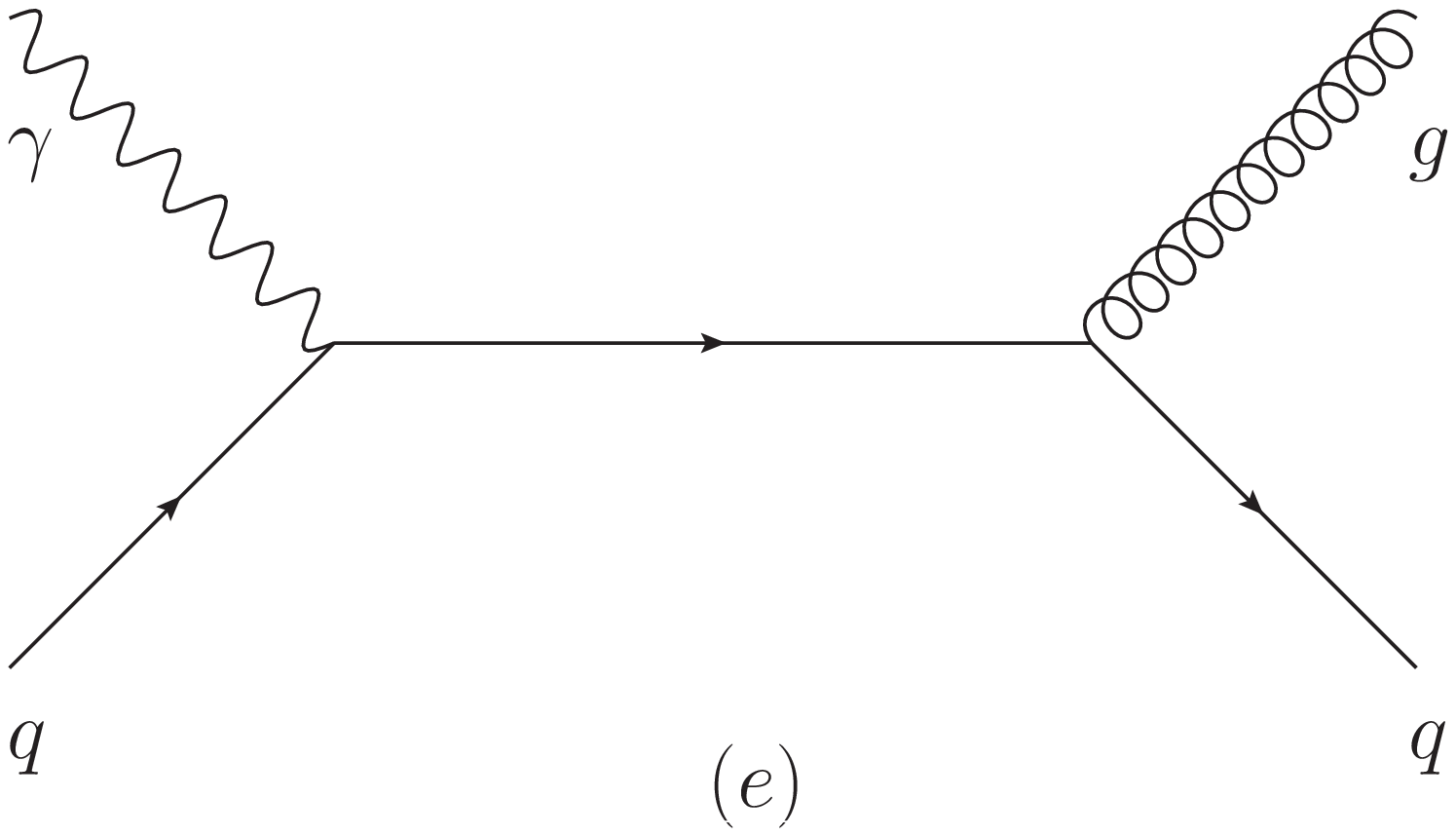}
  \end{minipage}
  \begin{minipage}[t]{0.3\linewidth}
  \centering
  \includegraphics[width=0.8\columnwidth]{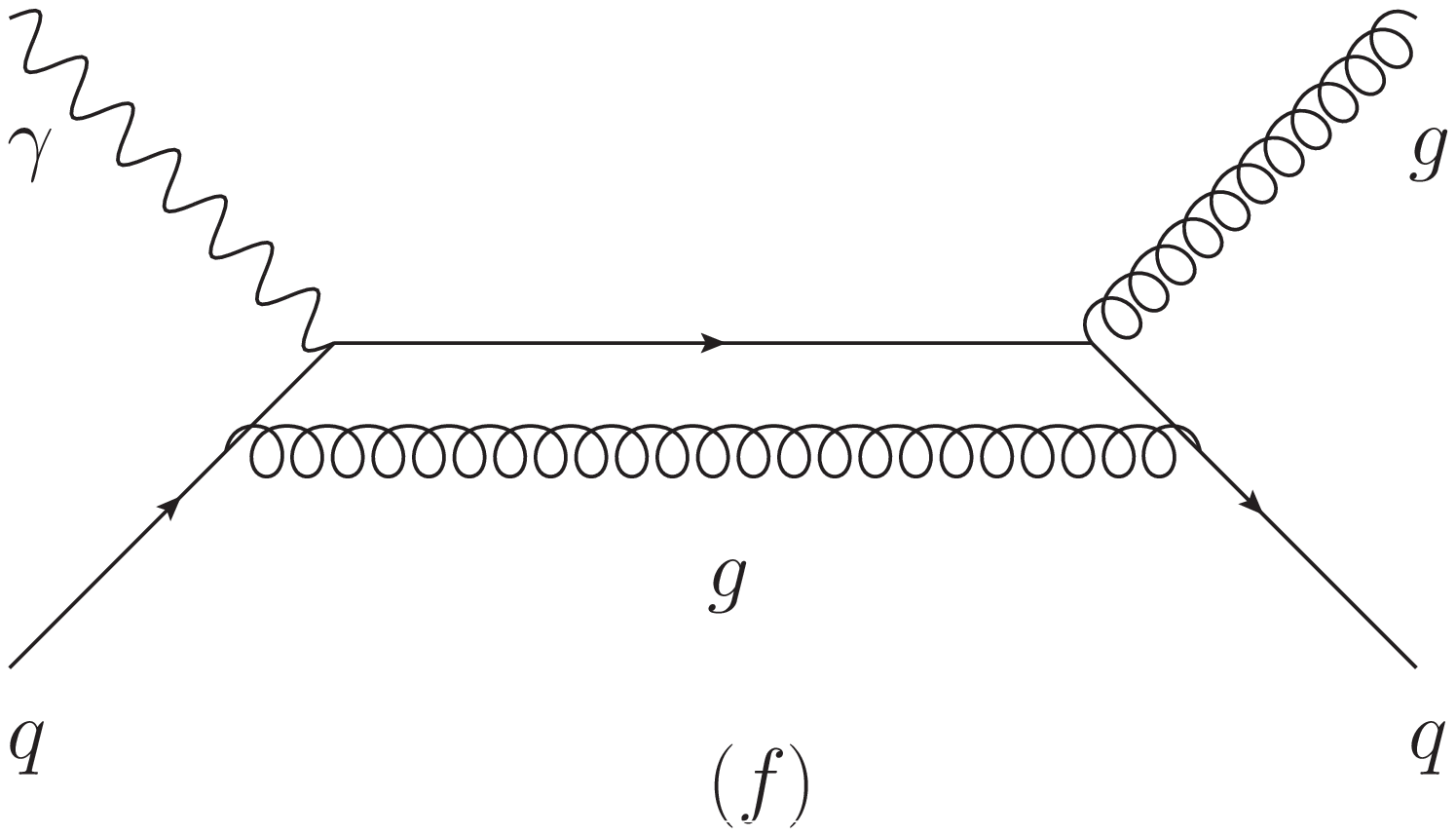}
  \end{minipage}
  \caption{Representative Feynman diagrams of the partonic subprocess in $\gamma^* +  p$  scattering which are included in the numerator (first line) and denominator (second line) of Eq.~\eqref{teec_with_pdf}. Here, the virtual photon is denoted by a wavy line and the gluon by a curled line.}\label{fig:fyenN}
\end{figure}

In the Breit frame,
the transverse energy-energy correlation in $\gamma(q) + p \to  a + b + X$ involving hadrons or jets is expressed  as:
\begin{align}
    \frac{1}{\sigma^\prime}\frac{d\Sigma^\prime}{d\cos\phi}{}&\equiv   \frac{\sum_{a,b} \int  dE_Td\cos\phi_{ab} \frac{d\sigma_{\gamma p\to   a+b+X}}{dE_Td\cos\phi_{ab}} \frac{2 E_{T,a} E_{T,b}}{|\sum_{i}E_{T,i}|^2} \delta(\cos \phi_{ab} - \cos \phi)}{ \int  dE_T d\sigma_{\gamma p\to a+b+X}/dE_T}\notag\\
    {}&=\frac{1}{N}\sum\limits^N_{A=1}\frac{1}{\Delta\cos\phi}\sum\limits_{pairs\, in \,\Delta\cos\phi}\frac{2E_{Ta}^AE^A_{Tb}}{(E_T^A)^2},
    \label{teec}
\end{align}
where $E_{T,a}$ and $E_{T,b}$ are transverse energies of two jets or hadrons.  The $\delta$-function assures that these hadrons
or jets are separated by the azimuthal angle $\phi$, and the cross section $\sigma^\prime$ and $\Sigma^\prime$ indicate kinematic cuts
on the integrals, defined later.
The second expression is valid for a sample of $N$ hard-scattering multi-jet events, labelled by the index $A$.
 The associated asymmetry (ATEEC) is then defined as the asymmetry between the
  forward ($\cos \phi >0$) and backward ($\cos \phi <0$)  parts of the TEEC:
\begin{align}
    \frac{1}{\sigma^\prime}\frac{d\Sigma^{\prime asym}}{d\cos\phi}{}&\equiv \frac{1}{\sigma^\prime}\frac{d\Sigma^\prime}{d\cos\phi}|_\phi-\frac{1}{\sigma^\prime}\frac{d\Sigma^\prime}{d\cos\phi}|_{\pi-\phi}.
    \label{ateec}
\end{align}
Due to the factorization of  the amplitudes in QCD, the denominator of the first equation in Eq.~(\ref{teec})  $d\sigma_{\gamma p\to   a+b+X}/dE_T$ can be written as a convolution of the parton distribution functions(PDFs) $f_{q/p}(x_1)$, where $x_1$ is the fractional energy of the proton carried by the parton $q$,  and the parton level
cross section. In the leading order, this is given by  $\sigma_{\gamma q\to b_1 b_2 }$.  As  for the numerator, it can also be expressed as the convolution of PDFs with $2\to 3$ parton level subprocess, in the leading order, such as $ \gamma q\to qgg$. Thus, TEEC is calculated from the following expression:
\begin{align}
    \frac{1}{\sigma^\prime}\frac{d\Sigma^\prime}{d\cos\phi} = \frac{ \sum_{j, a,b} \int  dE_T d\cos\phi_{ab} f_{j/p}(x_1)\star
    d\sigma_{\gamma j\to b_1 b_2 b_3}/(dE_Td\cos\phi_{ab})\frac{2 E_{T,a} E_{T,b}  \delta(\cos \phi - \cos \phi_{ab})}{|\sum_{i}E_{T,i}|^2}
    }{\sum_{j}f_{j/p}(x_1)\star\sigma_{\gamma j\to b_1 b_2 }},
    \label{teec_with_pdf}
\end{align}
where the symbol $\star$ stands for the convolution and j represents quarks and gluons. Which processes are included in the calculations of the TEEC depends on the
theoretical accuracy. In  NLO, this involves $2 \to 2$, $2 \to 3$ and $2 \to 4$ partonic subprocesses.
 Some representative Feynman diagrams of the subprocess are shown in Fig.~\ref{fig:fyenN}. In the upper row of Fig.~\ref{fig:fyenN}, we show the leading order (LO)  $(a)$, NLO real  $(b)$ and NLO virtual diagrams $(c)$ which enter in the calculations of the numerator of Eq.~\eqref{teec_with_pdf}. In the lower row of this figure, the Feynman diagrams of the subprocess in the denominator
 of  Eq.~\eqref{teec_with_pdf} are shown.  Of these, $(d,e)$ are LO diagrams,  and the NLO virtual corrections are represented by
 the diagram $(f)$. The NLO real diagram in inclusive two jet cross section are the same as the LO diagrams of three jet cross section of which we have shown a representative  diagram $(a)$ in Fig.~\ref{fig:fyenN}.

As defined in Eq.~\eqref{teec_with_pdf}, the TEEC correlation $\frac{1}{\sigma^\prime}\frac{d\Sigma^\prime}{d\cos\phi}{}$ is a normalized
variable. In particular,  the dependence of the TEEC on the PDFs is compensated to a large extent. Thus, to a good
approximation,  a factorized result is expected,
\begin{align}
    \frac{1}{\sigma^\prime}\frac{d\Sigma^\prime}{d\cos\phi} \sim \frac{\alpha_s(\mu)}{\pi} F(\cos\phi),
    \label{teec_with_alphas}
\end{align}
which can be perturbatively improved by including higher orders.

 We calculate the TEEC and ATEEC  close to experimental conditions used by the HERA experiments H1 and ZEUS,
 which assume  a certain selection criteria based on physical cuts on the kinematic variables. They are defined as follows:
The basic DIS kinematic variables $x$ and $y=Q^2/(s x)$ satisfy
 \begin{eqnarray}
{}&0<x<1,\,\,0.2<y<0.6 .
\end{eqnarray}
Besides, we restrict the range of the pseudo-rapidity in the laboratory frame ($\eta^{\text{lab}}$)  as
\begin{eqnarray}
-1<\eta^{\text{lab}}<2.5.
\end{eqnarray}
The pseudorapidity is related to the polar angle $\theta$, defined with respect to the proton beam direction, by $\eta^{\text{lab}}= - \ln \tan (\theta/2)$. We also use the right-handed co-ordinate system of the H1 collaboration, in which the positive $z$-axis is in the direction of the proton
 beam, and the nominal interaction point is located at $z=0$.

We calculate the TEEC and ATEEC in the Breit frame used by experiments at HERA.  In this frame, transverse energy $E_T$ is z-axis boost invariant and  $\gamma p\to jjjX$ is  the nontrivial process at the leading order.
The cuts for the transverse energy of dijet and trijet events, defined in the Breit frame,  are  as follows:
\begin{eqnarray}
\label{eq:etcuts}
  5\,\text{GeV} < \langle E_T\rangle_2 < 50 \,\rm{GeV}\notag\\
  5.5\,\text{GeV} < \langle E_T\rangle_3 < 40 \,\rm{GeV},
\end{eqnarray}
where the $\langle E_T\rangle_2$ and $\langle E_T\rangle_3$ denote $\frac{1}{2}({E}_T^{\text{jet}1}+{E}_T^{\text{jet}2})$ and $\frac{1}{3}({E}_T^{\text{jet}1}+{E}_T^{\text{jet}2}+{E}_T^{\text{jet}3})$, respectively. These cuts are consistent with the measurement at HERA \cite{Andreev:2016tgi}.
Following the practice in the HERA experimental analysis,
 we use the $k_T$ jet-algorithm~\cite{Ellis:1993tq}, where  the distance measure of partons ($i, j$) is given by
\begin{eqnarray}
{}&d_{ij}=\min(k_{ti}^{2},k_{tj}^{2})\frac{(\eta_i-\eta_j)^2+(\phi_i-\phi_j)^2}{R^2},\notag\\
{}&d_{iB}=k_{ti}^{2}.
\end{eqnarray}
Here $B$ represents the "beam jet" of the proton: particles with small momenta transverse to the beam axis, and
 $R$ is the cone-size  parameter of the jet which we set to $R=1.0$ in our calculation.
We  use two different PDF sets,  CT18~\cite{Hou:2019efy} and MMHT14~\cite{Harland-Lang:2014zoa},  and explore the uncertainty on the TEEC $(\cos \phi)$ and ATEEC $(\cos \phi)$ distributions  from these two sets in the next section.

It has become customary to determine the QCD coupling constant at the scale $\mu=M_Z$~\cite{Zyla:2020zbs}.
To  determine $\alpha_s(M_Z^2)$ from TEEC $(\cos \phi)$  and   ATEEC $(\cos \phi)$ in DIS,
 the cross section can be expressed as:
\begin{eqnarray}
\label{eq:sigmad}
\sigma=\sum_k \int dx f_k(x,\mu_F) \sigma_k(x,\mu_F,\mu_R),
\end{eqnarray}
where $k$ denotes a parton (quark or gluon), $ f_k(x, \mu_F)$ is the parton density,  and  $\sigma_k(x,\mu_F,\mu_R)$ is the partonic cross section, which  depends on the renormalization scale $\mu_R$ and the fatorization scale $\mu_F$. The partonic cross section is calculated in perturbative QCD as an expansion in $\alpha_s$:
\begin{eqnarray}
\sigma_k=\sum_n\alpha_s^n(\mu_R) \sigma^{(n)}_k(x,\mu_R,\mu_F).
\end{eqnarray}
As the $\mu_F$-dependence is very mild on
TEEC, as  shown later, the dominant scale-dependence of the cross section enters through the  scale $\mu_R$,  i.e., from $\alpha_s(\mu_R)$,
which we relate to  $\alpha_s(M_Z^2)$  on an event-by-event basis in our simulations. The $\mu_R$ dependence of $\alpha_s$  is given by renormalization group equation
\begin{eqnarray}
\mu_R\frac{d\alpha_s}{d\mu_R}=\beta(\alpha_s).
\end{eqnarray}
In the NLO calculation, the two-loop $\beta$-function is used for transcribing $\alpha_s(\mu)$ to $\alpha_s(M_Z^2)$ with a certain scale $\mu$ which is revelent for the jets defined above. The coupling constant $\alpha_s(\mu)$ is given as
\begin{eqnarray}\label{alphas}
\alpha_s(\mu)=\frac{1}{b_0\log(\mu^2/\Lambda^2)}\left[ 1-\frac{b_1\log(\log(\mu^2/\Lambda^2))}{b_0^2\log(\mu^2/\Lambda^2)}\right],\,b_0=\frac{33-2n_f}{12\pi},\,b_1=\frac{153-19n_f}{24\pi^2}.
\end{eqnarray}
Here, $n_f$ is the number of quark flavors, which is determined by the scale $\mu$, we have set $n_f=5$, and
 $\Lambda$ is the QCD parameter, which is determined by the value of $\alpha_s(M_Z^2)$. In the LO calculation, we set $b_1=0$ in
 the above expression. In our numerical results, we present TEEC$(\cos \phi)$ and ATEEC$(\cos \phi)$ calculated in the LO and NLO
 for the same value of $\alpha_s(M_Z^2)$, which implies a different value of $\Lambda$ in the LO and NLO.
  As already stated,  the $\alpha_s(\mu)$-dependence enters
 essentially through $\mu=\mu_R$. Since $\mu_R$ is not determined uiquely, there will remain a residual scale-dependence in the
 differential distributions for  TEEC $(\cos \phi)$  and   ATEEC $(\cos \phi)$.
In the next section, we show the dependence of TEEC $(\cos \phi)$  and   ATEEC $(\cos \phi)$ at HERA on the scales $\mu_F$, $\mu_R$
and $\alpha_s(M_Z^2)$.

Before ending this section, we remark that very recently another shape variable involving the azimuthal angle correlation of the
lepton and hadron in DIS process has been proposed and  calculated in~\cite{Li:2020bub}, which is defined as
\begin{align}
\ell HTEC(\cos \phi) \equiv &\Sigma_a \int d\sigma_{\ell p \to \ell + a +X} \frac{E_{T,\ell} E_{T,a}} {E_{T,\ell} \Sigma_i E_{T,i}}
 \delta(\cos \phi_{la} - \cos \phi)\notag\\
    {}&=\Sigma_a \int d\sigma_{\ell p \to \ell + a +X} \frac{E_{T,a}} {\Sigma_i E_{T,i}}
 \delta(\cos \phi_{la} - \cos \phi),
\label{eq:lhtec}
\end{align}
where the sum runs over all hadrons and $\cos \phi_{\ell a}$ is the cosine of the azimuthal angle between the lepton and the
hadron. As seen in the second of the above equation, transverse energy of the lepton drops out of this variable.
 As opposed to the
shape variable TEEC, defined here in Eq.~(\ref{teec})  for DIS, as well as the  EEC/TEEC variables defined earlier in
$e^+e^-$ annihilation~\cite{Basham:1978bw,Basham:1978zq} and $pp$ collisions~\cite{Ali:1984yp}, which involve (transverse) energy weighted azimutal angle correlations between two jets
or hadrons, the shape variable defined in~\cite{Li:2020bub} is the azimuthal angle correlation between the lepton and a hadron (or a jet)
weighted by the transverse energy of a single  hadron (or jet). We emphasize that $\ell HTEC(\cos \phi)$, defined in~\cite{Li:2020bub}
and Eq.~(\ref{eq:lhtec}), while interesting in  its own right,  is a different variable  from TEEC.
Lepton-jet correlation in DIS has also been studied in~\cite{Liu:2018trl}, and revisited very recently in~\cite{Liu:2020dct}, where a detailed
derivation of the formalism used and a phenomenological study relevant for the jet production at HERA are carried out.

\section{Results for TEEC $(\cos \phi)$  and its asymmetry ATEEC  $(\cos \phi)$ in DIS process at HERA}

For the numerical results presented here in the LO and NLO accuracy, we have used the  program
 NLOJET++~\cite{Nagy:2001xb,Nagy:2001fj}. As a cross check on our calculations, we have also used the program  Madgraph to calculate the leading order TEEC and ATEEC functions. The errors shown for the TEEC and ATEEC are of
 statistical origin, resulting from the Monte Carlo phase space integration.
To compare with the results obtained using \textsc{NLOJET++}, parton-level events are generated in \text{MadGraph5\_aMC@NLO}~\cite{Alwall:2014hca}  with the MMHT14  PDF set.
  The distributions obtained from the two packages agree well in both the low-$Q^2$ ($5.5\,\rm{GeV}^2<Q^2<80\,\rm{GeV}^2$) and high-$Q^2$ ($150\,\rm{GeV}^2<Q^2<1000\,\rm{GeV}^2$) ranges. The details are given in Appendix A. From now on, we shall work only with the NLOJET++.

 We have generated $10^9$ events to obtain the LO and NLO results
 in each of the two $Q^2$ ranges. This large statistics is required to obtain an accuracy of a few percent in NLO,
 which enables us to meaningfully calculate the various parametric dependences intrinsic to the problem at hand.
At the very outset, we have  calculated the two-jet cross sections at $\sqrt{s}=314$ GeV for the ranges of the DIS variables given in the preceding section and
compared them with the corresponding HERA data~\cite{Andreev:2016tgi} in Table~I. The NLOJET++ results are obtained from the jet clusters using parton-level cross sections and the HERA data refer to the hadron-level cross section. While the two are not
identical, this comparison should hold to a good first approximation. The two-jet events selected for
this comparison are  defined by the following two bins in $\langle E_T\rangle_2 $   and  the $Q^2$-range given below:
\begin{eqnarray}
 {}&5.5\,\rm{GeV}^2 < Q^2< 8 \,\rm{GeV}^2,\notag\\
 {}&{\rm bin1}: 5\,\rm{GeV} < \langle E_T\rangle_2 < 7 \,\rm{GeV},\notag\\
 {}&{\rm bin2}: 7\,\rm{GeV} < \langle E_T\rangle_2 < 11 \,\rm{GeV}.
\end{eqnarray}
Theoretical  cross sections are obtained using the CT18~\cite{Hou:2019efy} PDFs, the scales set to the values $\mu_R=\mu_F=\sqrt{\langle E_T \rangle^2 +Q^2}$, and $\alpha_s(M_Z)=0.118$. The NLO  cross sections  are in excellent agreement
with the HERA data.

\begin{table}[htbp!]\label{tab:tab1}
	\centering
	\caption{Dijet cross sections at HERA with $\sqrt{s}=$314 GeV in the two $\langle E_T\rangle_2 $-bins defined in the text and the
	corresponding  HERA data  from the Table~7 in H1 collaboration~\cite{Andreev:2016tgi} . }
	\begin{tabular}{|l|c|c|}
		\hline
		&bin1&bin2 \\
		\hline
		$\sigma_{\text{HERA}}$[pb]& $299\pm9.9\pm52.3$&$185\pm 3.7 \pm 13.9$ \\
		\hline
		$\sigma_{\text{NLOJET++}}$[pb]& $298.03\pm 3.93$& $199.9\pm3.04$\\
		\hline
	\end{tabular}
\end{table}

\begin{figure}[htbp!]
   \begin{minipage}[t]{0.45\linewidth}
  \centering
  \includegraphics[width=1.0\columnwidth]{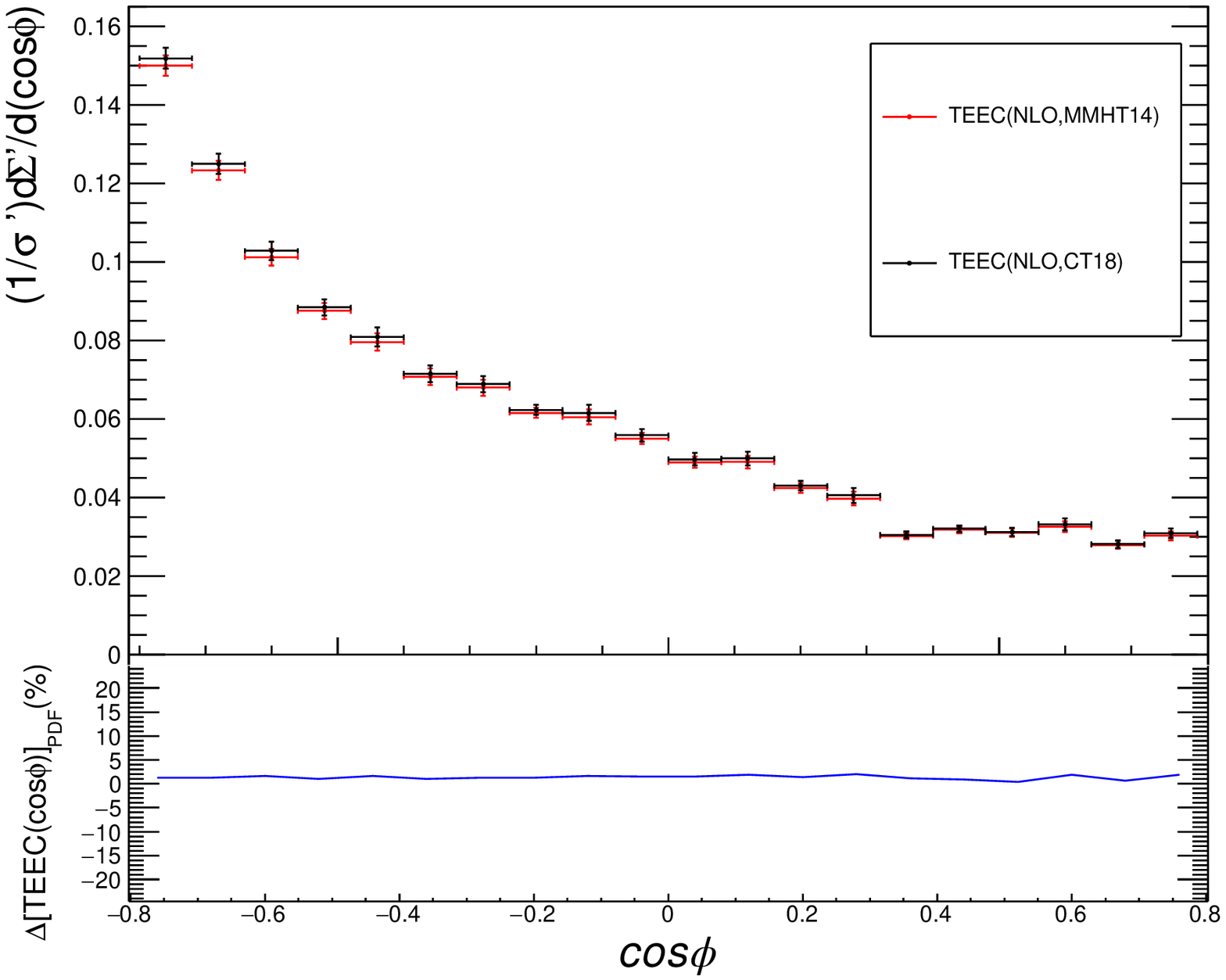}
  \end{minipage}
   \begin{minipage}[t]{0.45\linewidth}
  \centering
  \includegraphics[width=1.0\columnwidth]{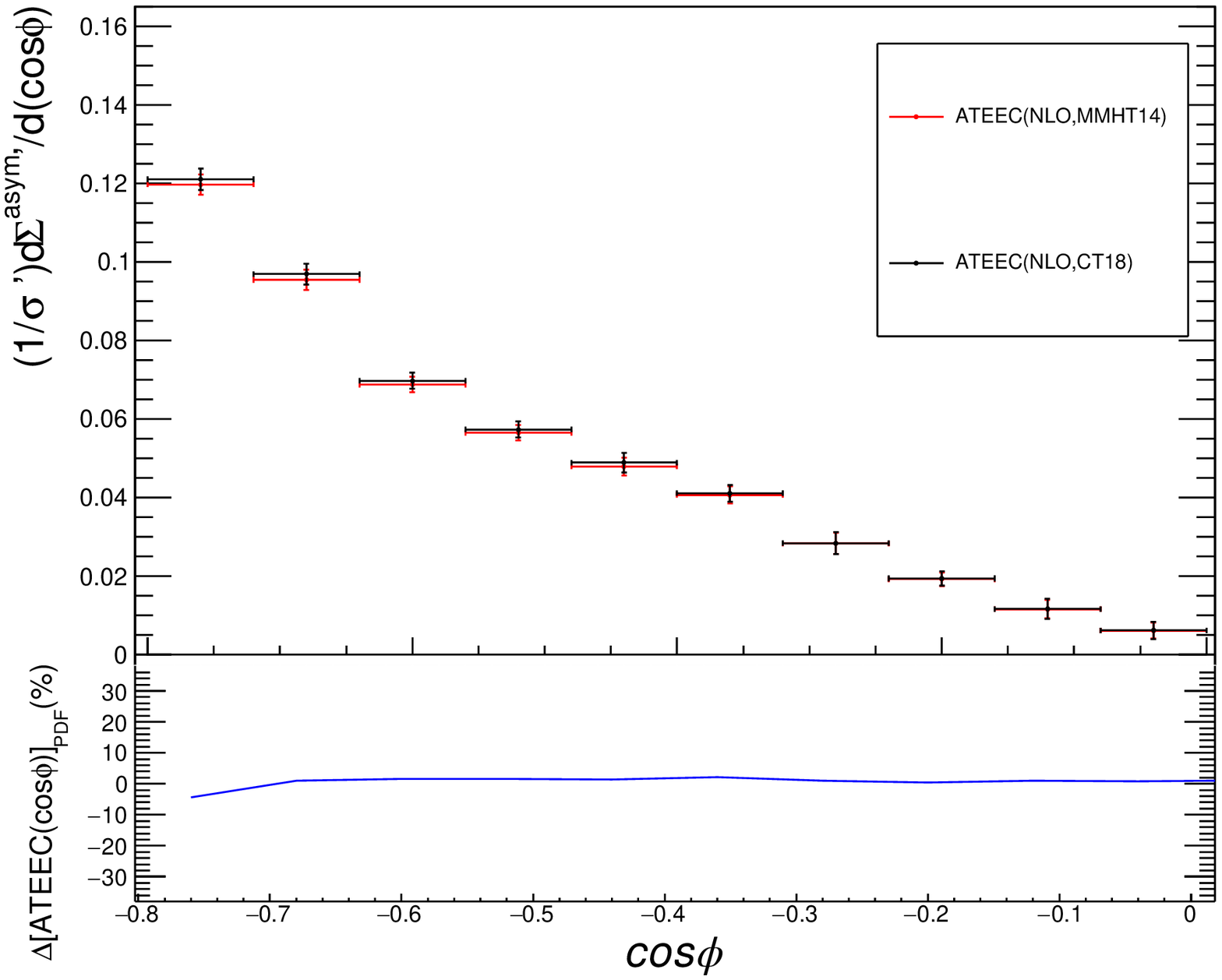}
  \end{minipage}
  \caption{Differential distribution $1/\sigma^\prime d\Sigma^\prime/d (\cos \phi)$  and  its asymmetry
   $ 1/\sigma^\prime d\Sigma^{\prime asym}/d(\cos\phi) $,
  calculated  in Next-to-Leading order for the low-$Q^2$ range $5.5\,\rm{GeV}^2<Q^2<80\,\rm{GeV}^2$ for the $ep$ center-of-mass energy $\sqrt{s}= 314$ GeV at HERA. The two input PDFs are indicated on the upper frames. The lower frames show
   $\Delta [\text{TEEC}(\cos \phi)]_{\text{PDF}} $ and $\Delta [\text{ATEEC}(\cos \phi)]_{\text{PDF}}$, defined in Eq.~(\ref{eq:dpdf}).}
 \label{pdflow}
\end{figure}

\begin{figure}[htbp!]
   \begin{minipage}[t]{0.40\linewidth}
  \centering
  \includegraphics[width=1.0\columnwidth]{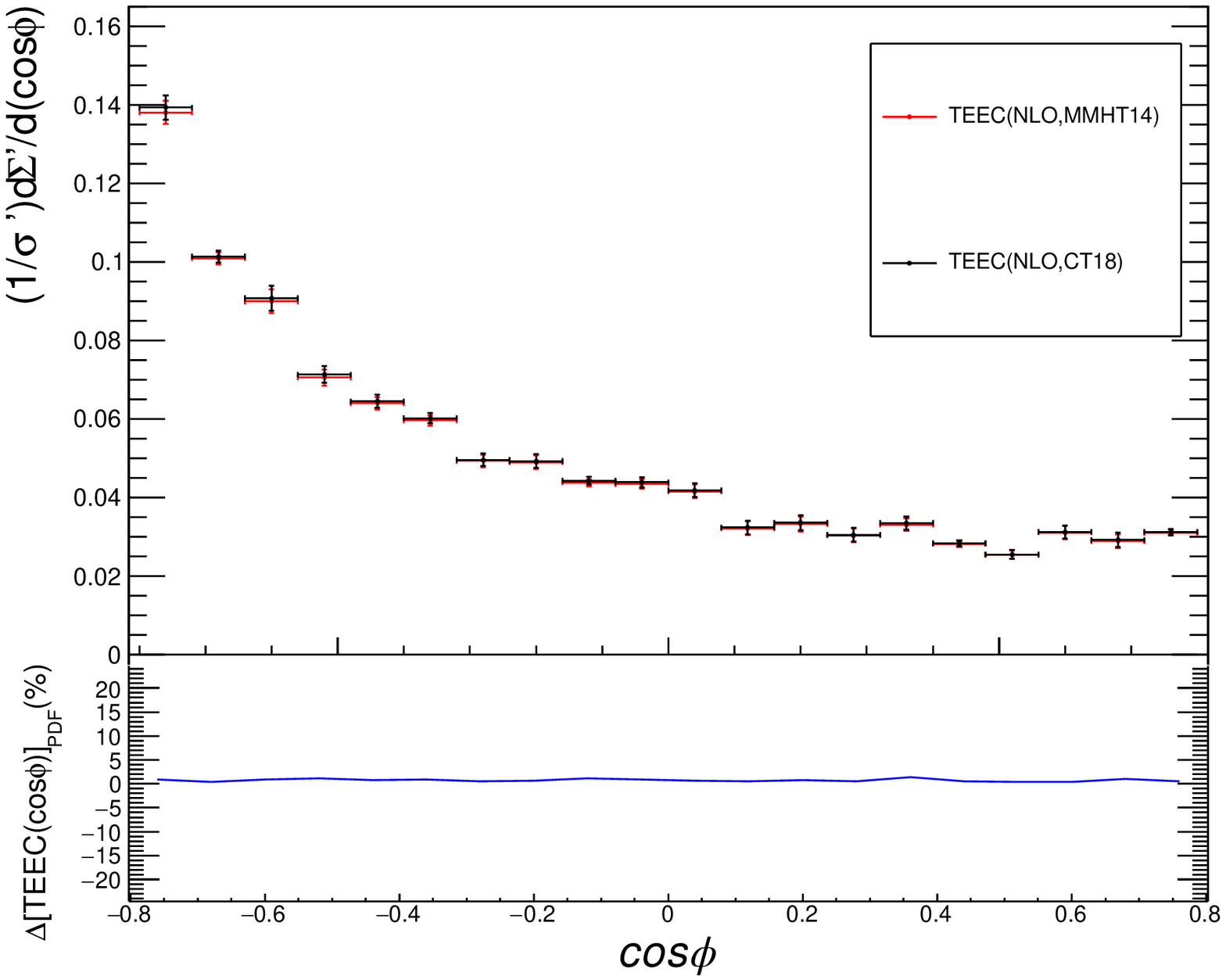}
  \end{minipage}
   \begin{minipage}[t]{0.40\linewidth}
  \centering
  \includegraphics[width=1.0\columnwidth]{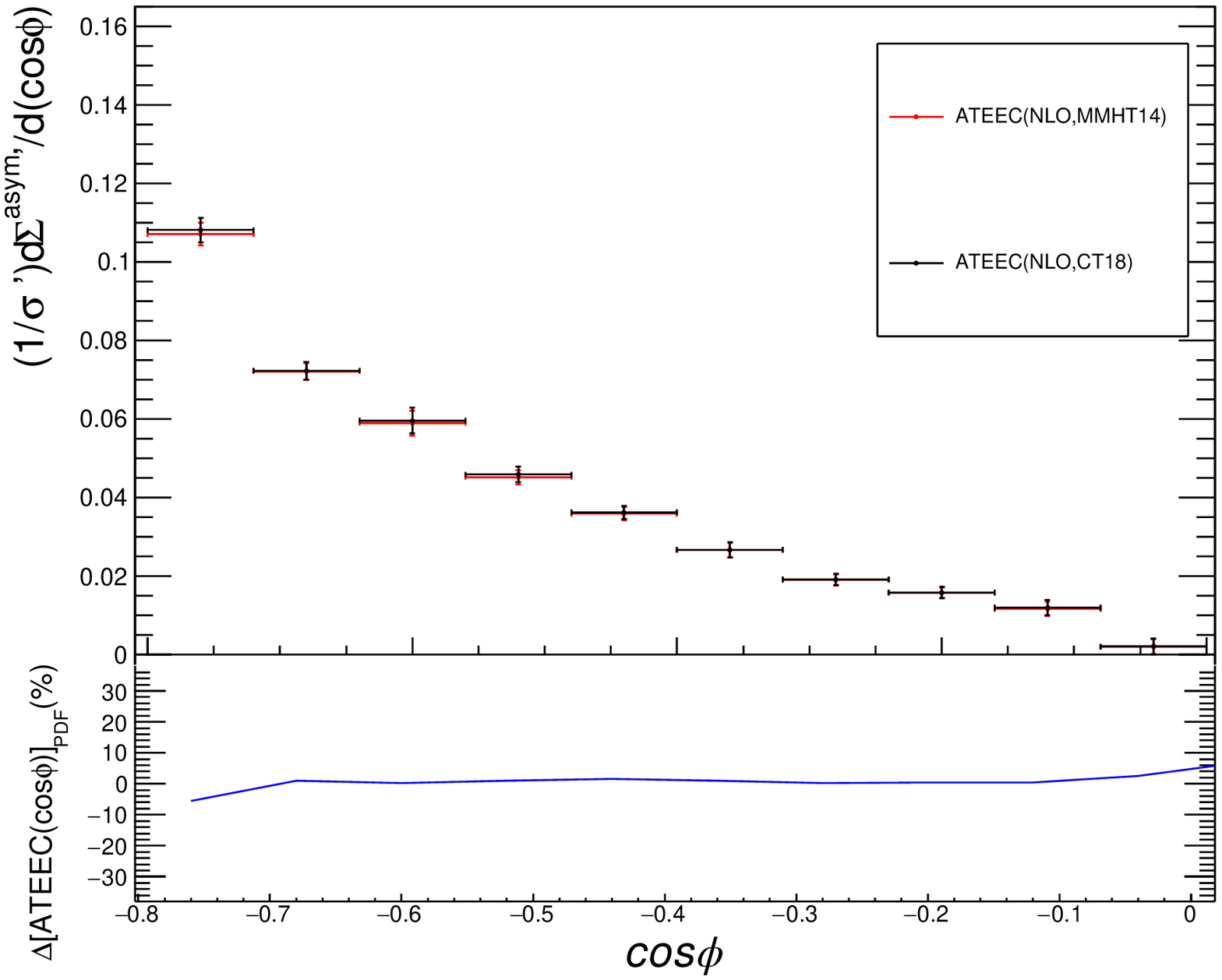}
  \end{minipage}
  \caption{Differential distribution $1/\sigma^\prime d\Sigma^\prime/d (\cos \phi)$  and  its asymmetry
   $ 1/\sigma^\prime d\Sigma^{\prime asym}/d(\cos\phi) $ as in Fig.~\ref{pdflow}, but for the high-$Q^2$ range $150\,\rm{GeV}^2<Q^2<1000\,\rm{GeV}^2$ at HERA. }\label{pdfhigh}
\end{figure}

We start by showing  the differential distributions $\frac{1}{\sigma^\prime}\frac{d\Sigma^\prime}{d\cos\phi}$, defining
TEEC $(\cos \phi)$,  and its asymmetry, $ \frac{1}{\sigma^\prime}\frac{d\Sigma^{\prime asym}}{d\cos\phi}, $  ATEEC $(\cos \phi)$, for the two PDF sets CT18~\cite{Hou:2019efy}
and MMHT14~\cite{Harland-Lang:2014zoa}. They are presented for the
 high-$Q^2$ range (50 $\rm GeV^2 ~\leq1000~\rm GeV^2$)
and the low-$Q^2$ range (5.5 $\rm{GeV}^2 \leq 80~\rm{GeV}^2$) in Fig.~\ref{pdflow} and Fig.~\ref{pdfhigh}, respectively. The left frame in these figure shows TEEC $(\cos\phi)$ and the right frame ATEEC $(\cos\phi)$, calculated in the NLO accuracy.

We restrict $\cos \phi$ in the range $[-0.8,0.8]$ to avoid the regions $\phi\simeq 0^\circ$ and $\phi \simeq  180^\circ$ which will involve self-correlations $(a=b)$ and virtual corrections to $2\to 2$ processes.
 In calculating these functions, we use $\alpha_s(M_Z)=0.118$ and have set the
fatorization ($\mu_F$) and the renormalization  ($\mu_R$) scales to the following values: $\mu_F=\mu_R=\mu_0=\sqrt{\langle E_T\rangle^2+Q^2}$. This scale-setting is discussed in the analysis of the jet-data by the H1 Collaboration~\cite{Andreev:2017vxu}. The effect of varying the scale $\mu_F$ which enters in the PDFs has  little effect in the inclusive- and dijet- cross sections~\cite{Andreev:2017vxu}, which we also  find for the TEEC $(\cos \phi)$ and ATEEC$(\cos \phi)$, shown later in this section.
 We quantify the uncertainty on the  TEEC $(\cos \phi)$ and ATEEC $(\cos \phi)$ from the two input PDFs by
the following ratios:
\begin{eqnarray}
\label{eq:dpdf}
&&  \Delta[\text{TEEC}(\cos \phi)]_{\rm pdf} \equiv \frac{\text{TEEC}(\cos \phi)_{\rm CT18}  - \text{TEEC}(\cos \phi)_{\rm MMHT14}} {\text{TEEC}(\cos \phi)_{\rm CT18} } \notag \\
&&
\Delta[\text{ATEEC}(\cos \phi)]_{\rm pdf} \equiv \frac{\text{ATEEC}(\cos \phi)_{\rm CT18}  - \text{ATEEC}(\cos \phi)_{\rm MMHT14}} {\text{ATEEC}(\cos \phi)_{\rm CT18} } .
 \end{eqnarray}
They are shown in the lower frames of Fig.~\ref{pdflow} and Fig.~\ref{pdfhigh}, respectively. The PDF-related uncertainties $ \Delta[\text{TEEC}(\cos \phi)]_{\rm pdf}$  and $ \Delta[\text{ATEEC}(\cos \phi)]_{\rm pdf} $ are shown in the lower frames in these figures. We note that, within our statistics,  the former are below
10\% for most of the range. The asymmetry becomes increasingly  small as one approaches  $\cos \phi \simeq 0$, and it would require much higher statistics to reduce the numerical error on  $ \Delta[\text{ATEEC}(\cos \phi)]_{\rm pdf} $ near the end-point. It should be stressed that when showing the ratios in the lower frames, the statistical errors have been neglected.

\begin{figure}[htbp!]
  \begin{minipage}[t]{0.40\linewidth}
  \centering
  \includegraphics[width=1.0\columnwidth]{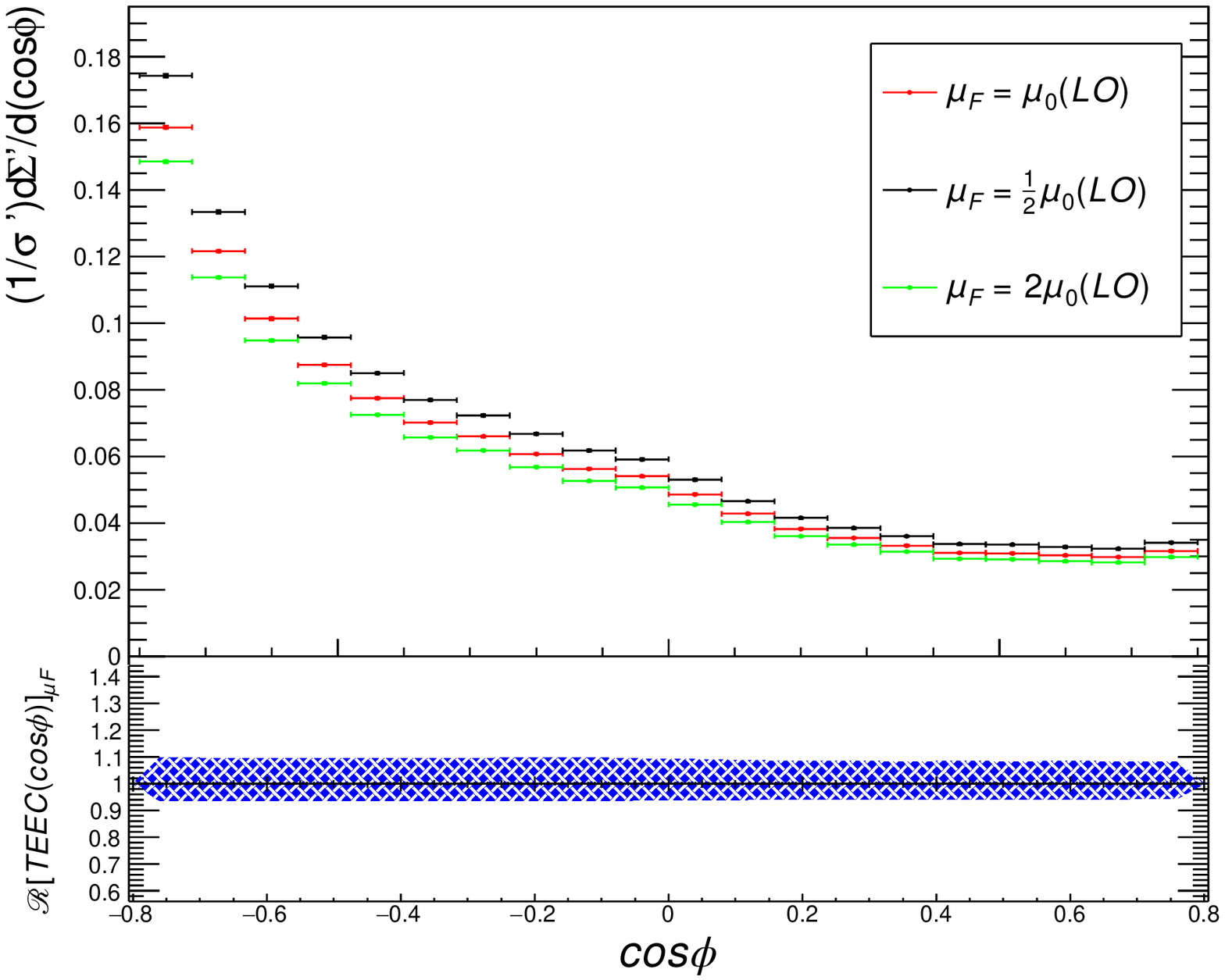}
    \end{minipage}
   \begin{minipage}[t]{0.40\linewidth}
  \centering
  \includegraphics[width=1.0\columnwidth]{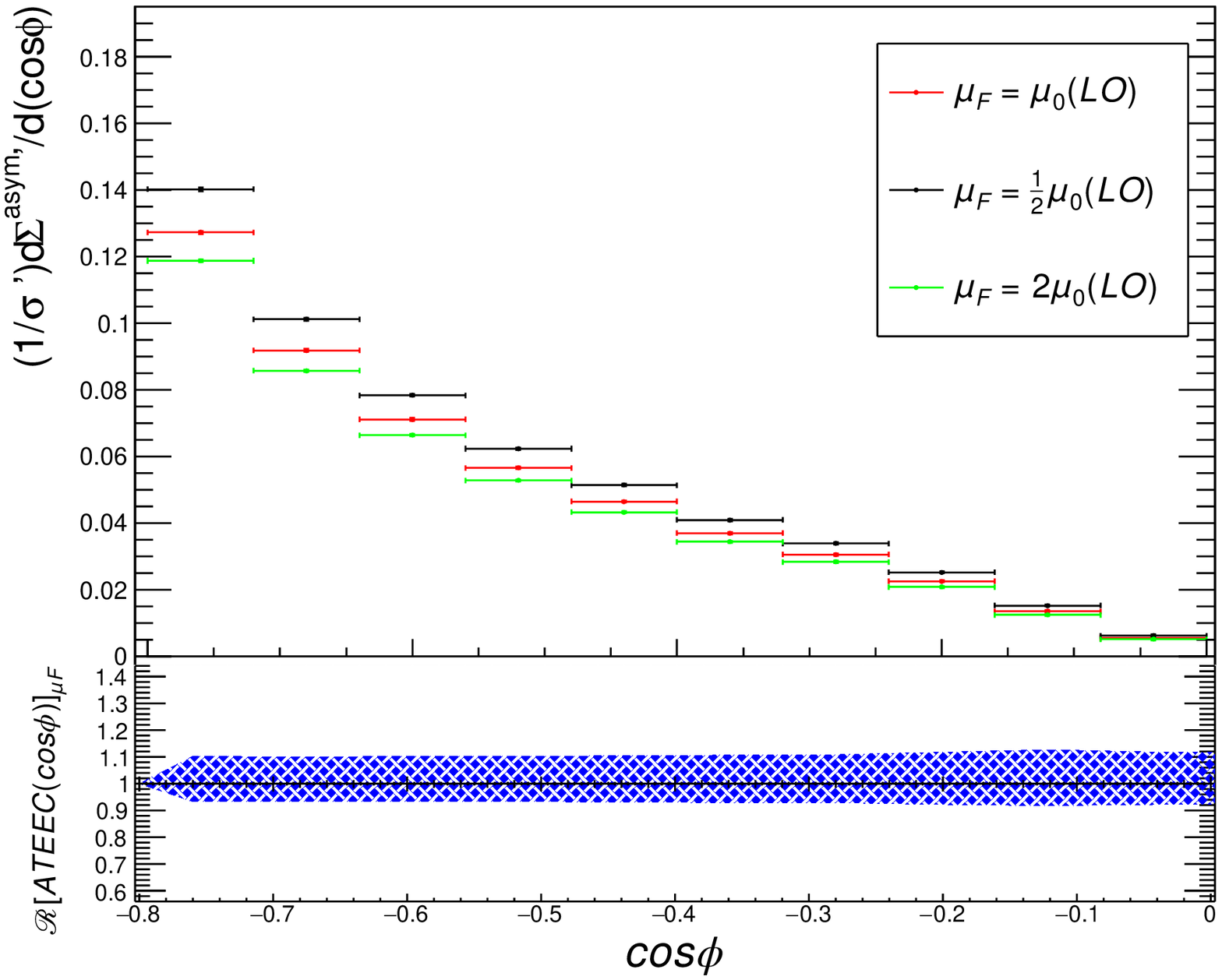}
  \end{minipage}
  \begin{minipage}[t]{0.40\linewidth}
  \centering
  \includegraphics[width=1.0\columnwidth]{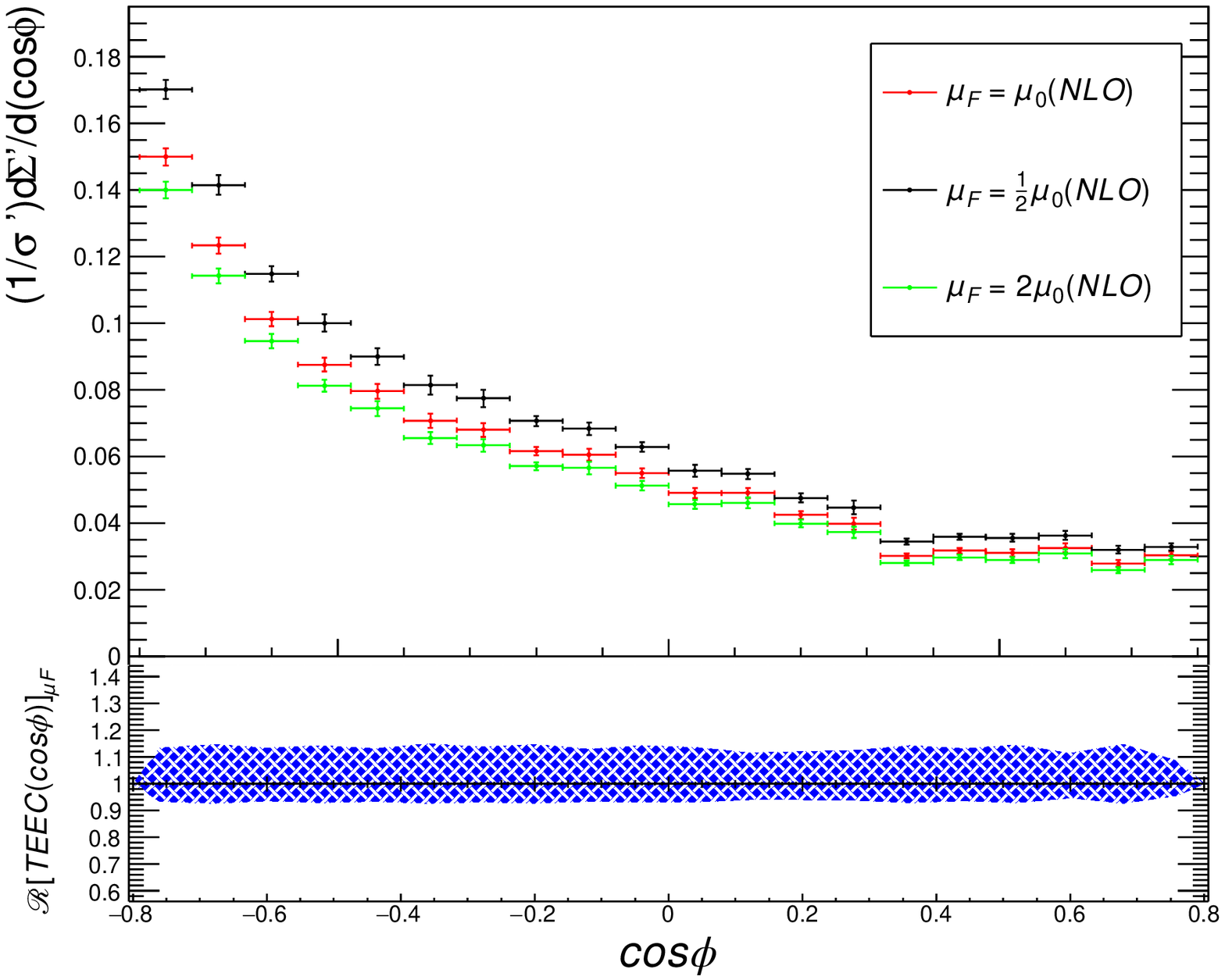}
    \end{minipage}
    \begin{minipage}[t]{0.40\linewidth}
  \centering
  \includegraphics[width=1.0\columnwidth]{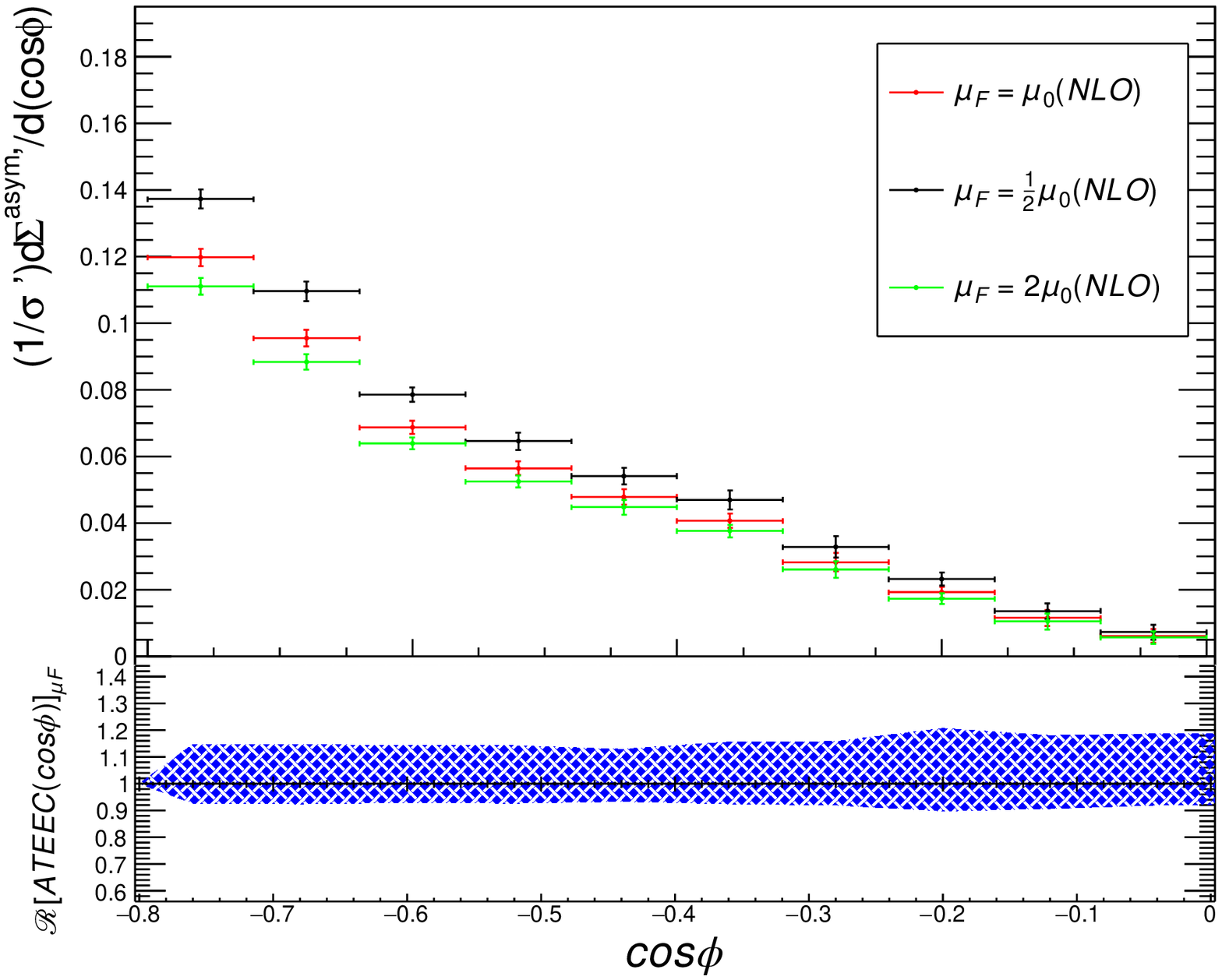}
    \end{minipage}

    \caption{Fatorization scale dependence  of the differential distribution $1/\sigma^\prime d\Sigma^\prime/d (\cos \phi)$  and  its asymmetry
   $ 1/\sigma^\prime d\Sigma^{\prime asym}/d(\cos\phi) $ in the leading order
(upper frames), and the next to leading order  (lower frames),  varying $\mu_F$ in the range $[0.5,2]\times \mu_0$, where
$\mu_0$ is the nominal scale defined in the text, calculated  with the
MMHT14 PDFs  for the low-$Q^2$: $5.5\,\rm{GeV}^2<Q^2<80\,\rm{GeV}^2$ at HERA.
 The corresponding $\mu_F$-dependence is also
shown in terms of ${\mathscr R} [{\rm TEEC} (\cos \phi)]_{\mu_F}$ and ${\mathscr R} [{\rm ATEEC} (\cos \phi)]_{\mu_F}$, defined in
Eq.~(\ref{eq:dmuF}).}
 \label{fig:fig4}
\end{figure}

\begin{figure}[htbp!]
  \begin{minipage}[t]{0.40\linewidth}
  \centering
  \includegraphics[width=1.0\columnwidth]{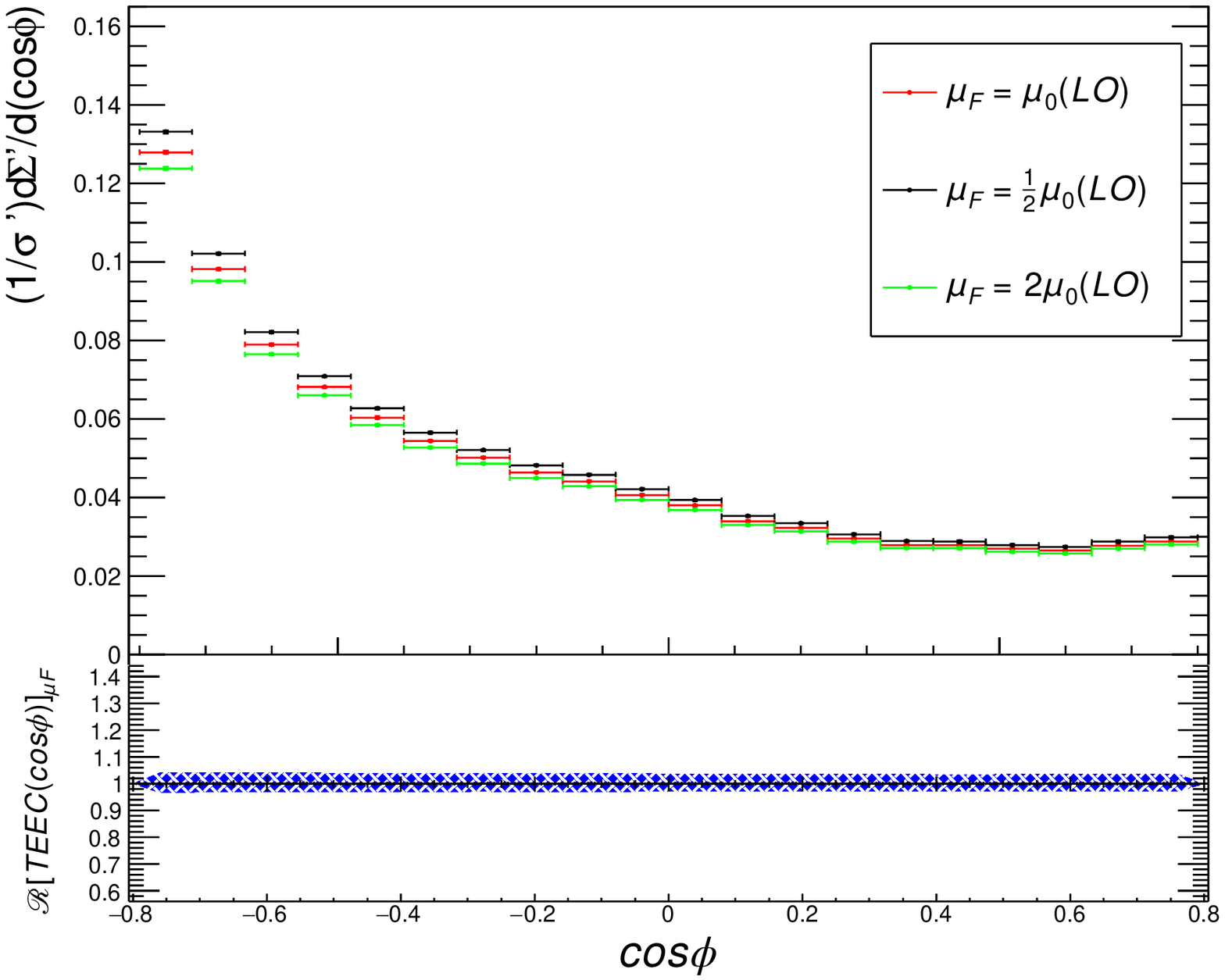}
    \end{minipage}
   \begin{minipage}[t]{0.40\linewidth}
  \centering
  \includegraphics[width=1.0\columnwidth]{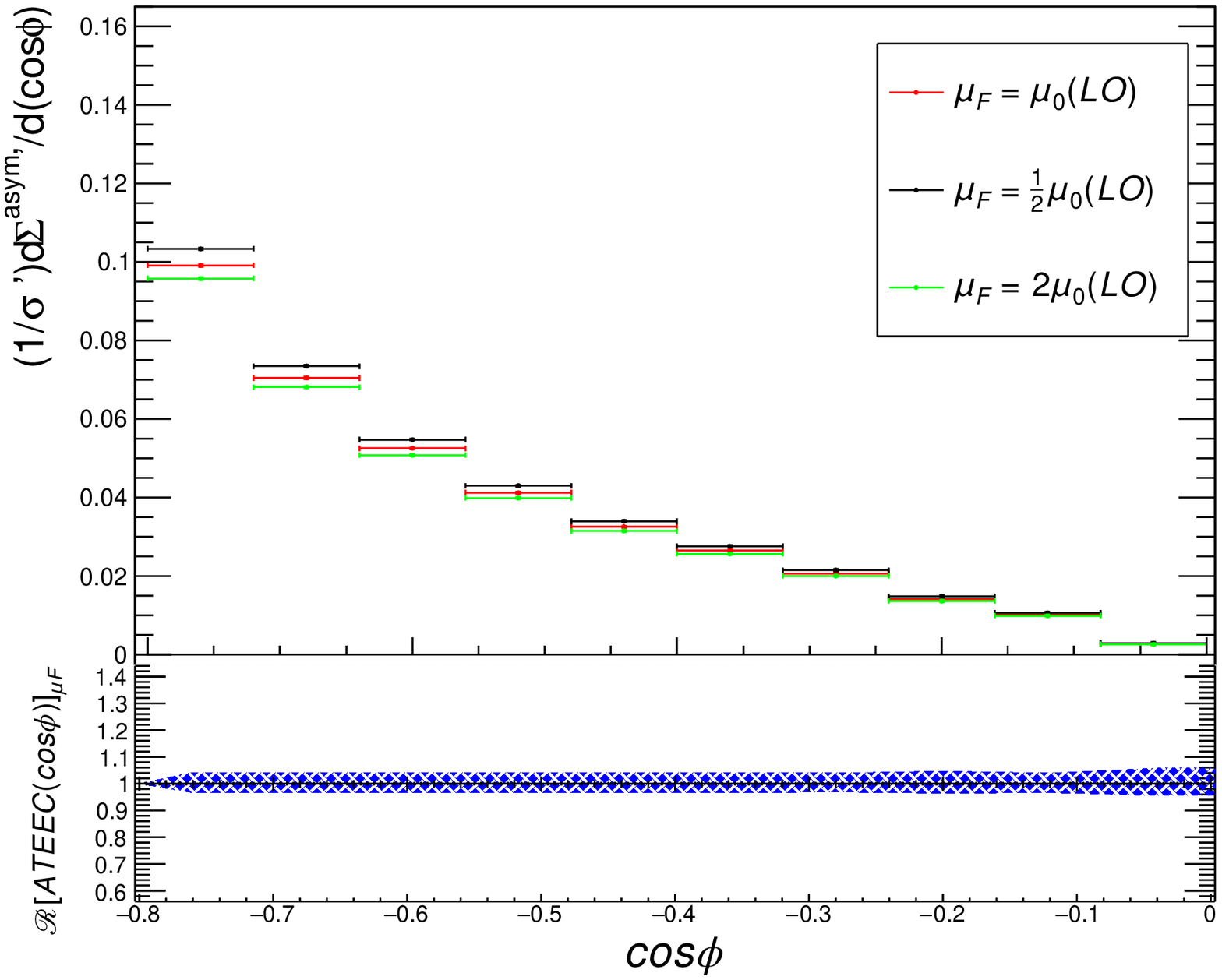}
  \end{minipage}
  \begin{minipage}[t]{0.40\linewidth}
  \centering
  \includegraphics[width=1.0\columnwidth]{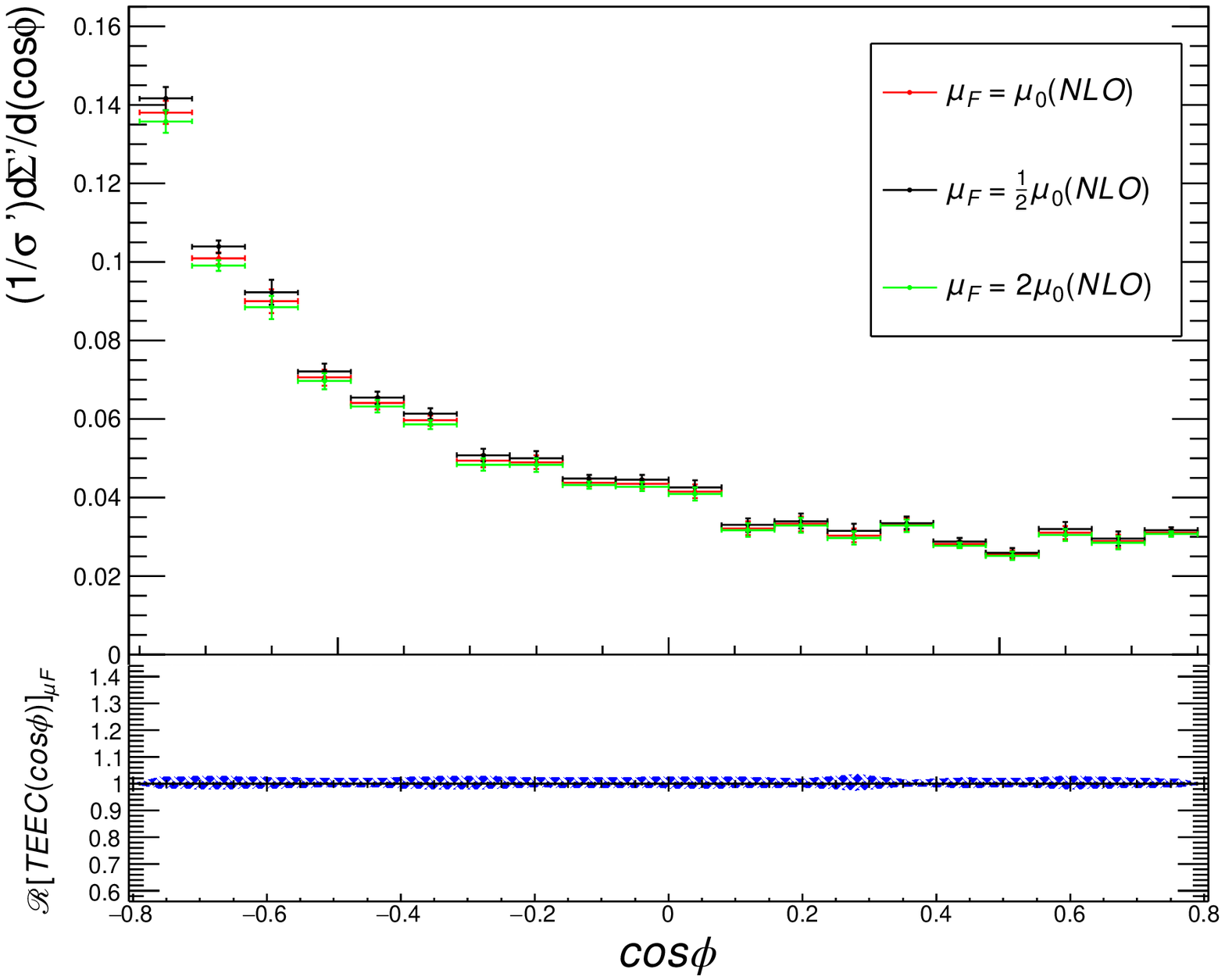}
    \end{minipage}
    \begin{minipage}[t]{0.40\linewidth}
  \centering
  \includegraphics[width=1.0\columnwidth]{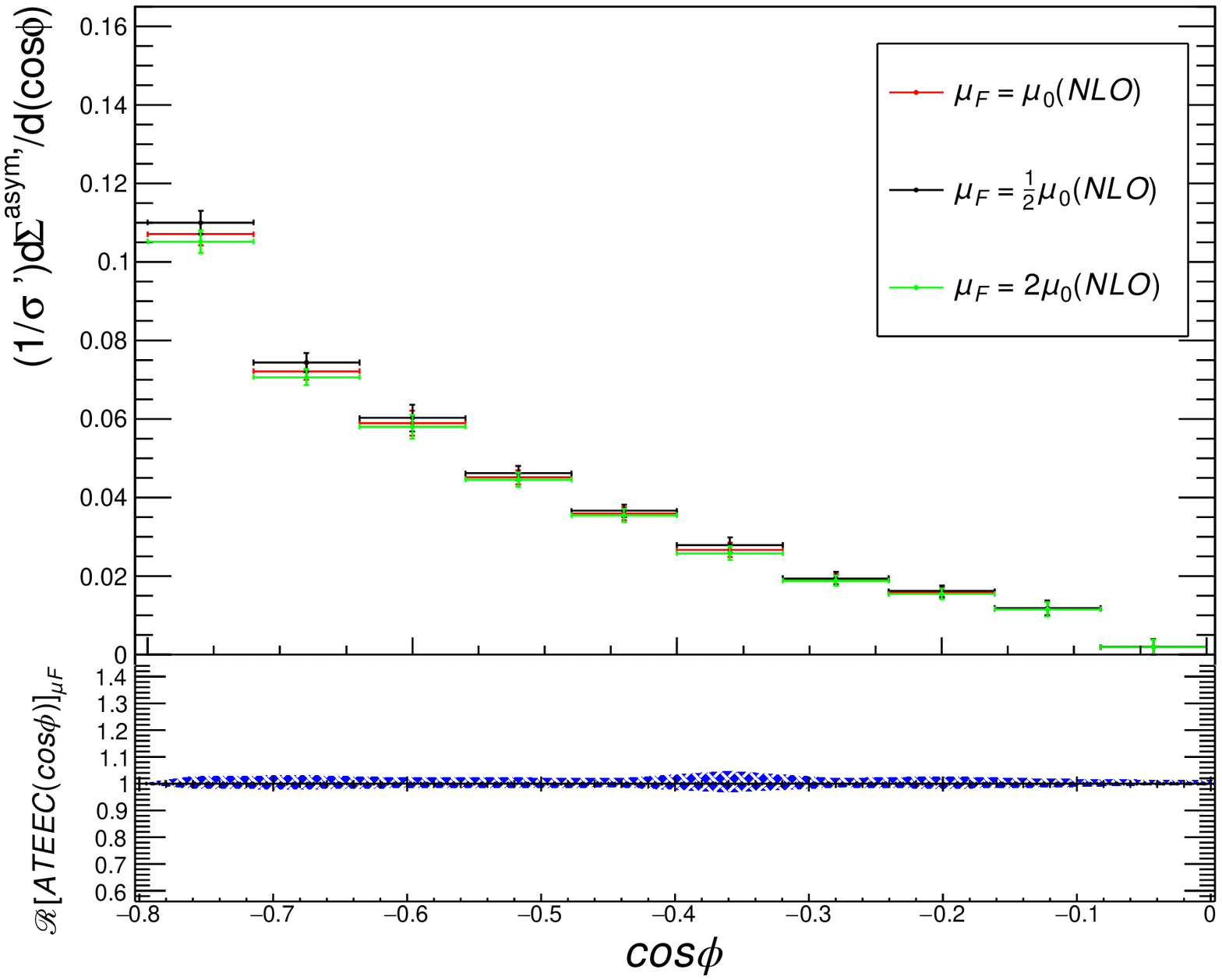}
    \end{minipage}

 \caption{Fatorization scale dependence  of the differential distribution $1/\sigma^\prime d\Sigma^\prime/d (\cos \phi)$  and  its asymmetry
   $ 1/\sigma^\prime d\Sigma^{\prime asym}/d(\cos\phi) $ in the leading order
 (upper frames) and the next to leading order (lower frames) as in Fig.~\ref{fig:fig4}, but for the high-$Q^2$ range
$150\,\rm{GeV}^2<Q^2<1000\,\rm{GeV}^2$ at HERA.}
\label{fig:fig5}
\end{figure}

Next, we  present
  the fatorization-scale and the renormalization-scale  dependence of the TEEC $(\cos \phi)$  and   ATEEC $(\cos \phi)$, by fixing the other parameters to their nominal values, and use the MMHT14 PDF set. Fixing $\mu_R=\mu_0$, we vary $\mu_F$ in the range
 $\mu_F=[0.5, 2]\mu_0$ and show  the  $\mu_F$-dependence in Fig.~\ref{fig:fig4}
and~Fig.~\ref{fig:fig5} for the low-$Q^2$ range $5.5\,\rm{GeV}^2<Q^2<80\,\rm{GeV}^2$ and the high-$Q^2$ range
 $150\,\rm{GeV}^2<Q^2<1000\,\rm{GeV}^2$, respectively, in  the LO and the NLO accuracy. The $\mu_F$-uncertainty on  TEEC $(\cos \phi)$ and ATEEC $(\cos \phi)$ are  plotted in the lower frames of
 these figures (Figs.~\ref{fig:fig4} and~\ref{fig:fig5})  in terms of the ratios $\mathscr{R}$[TEEC$(\cos \phi)]_{\mu_F}$
 and $\mathscr{R}$[ATEEC$(\cos \phi)]_{\mu_F}$ defined below
\begin{eqnarray}
\label{eq:dmuF}
&&  {\mathscr R}[\text{TEEC}(\cos \phi)]_{\mu_F} \equiv \frac{\text{TEEC}(\cos \phi)_{\mu_F=x \mu_0,\mu_R=\mu_0} } {\text{TEEC}(\cos \phi)_{\mu_R=\mu_F=\mu_0}},\quad x \in [0.5,2] , \notag \\
&&
{\mathscr R}[\text{ATEEC}(\cos \phi)]_{\mu_F} \equiv \frac{\text{ATEEC}(\cos \phi)_{\mu_F=x \mu_0,\mu_R=\mu_0} } {\text{ATEEC}(\cos \phi)_{\mu_R=\mu_F=\mu_0}},\quad x \in [0.5,2].
 \end{eqnarray}
 The $\mu_F$-dependence shown for $\mathscr{R}$[TEEC$(\cos \phi)]_{\mu_F}$ and $\mathscr{R}$[ATEEC$(\cos \phi)]_{\mu_F}$ is the
 resulting envelope by varying the scale $\mu_F$ in the indicated range and  the statistical errors arising from the numerical integration of the phase space. The statistical errors of TEEC $(\cos \phi)$  and   ATEEC $(\cos \phi)$ are about $3\% $.  The $\mu_F$-dependence of TEEC $(\cos \phi)$ is
 small,  decreasing for the high-$Q^2$ range. It is comparatively smaller for the asymmetry ATEEC $(\cos \phi)$, except for the
 last bin, where the numerical integration has  a large statistical error.

\begin{figure}[htbp!]
  \begin{minipage}[t]{0.40\linewidth}
  \centering
  \includegraphics[width=1.0\columnwidth]{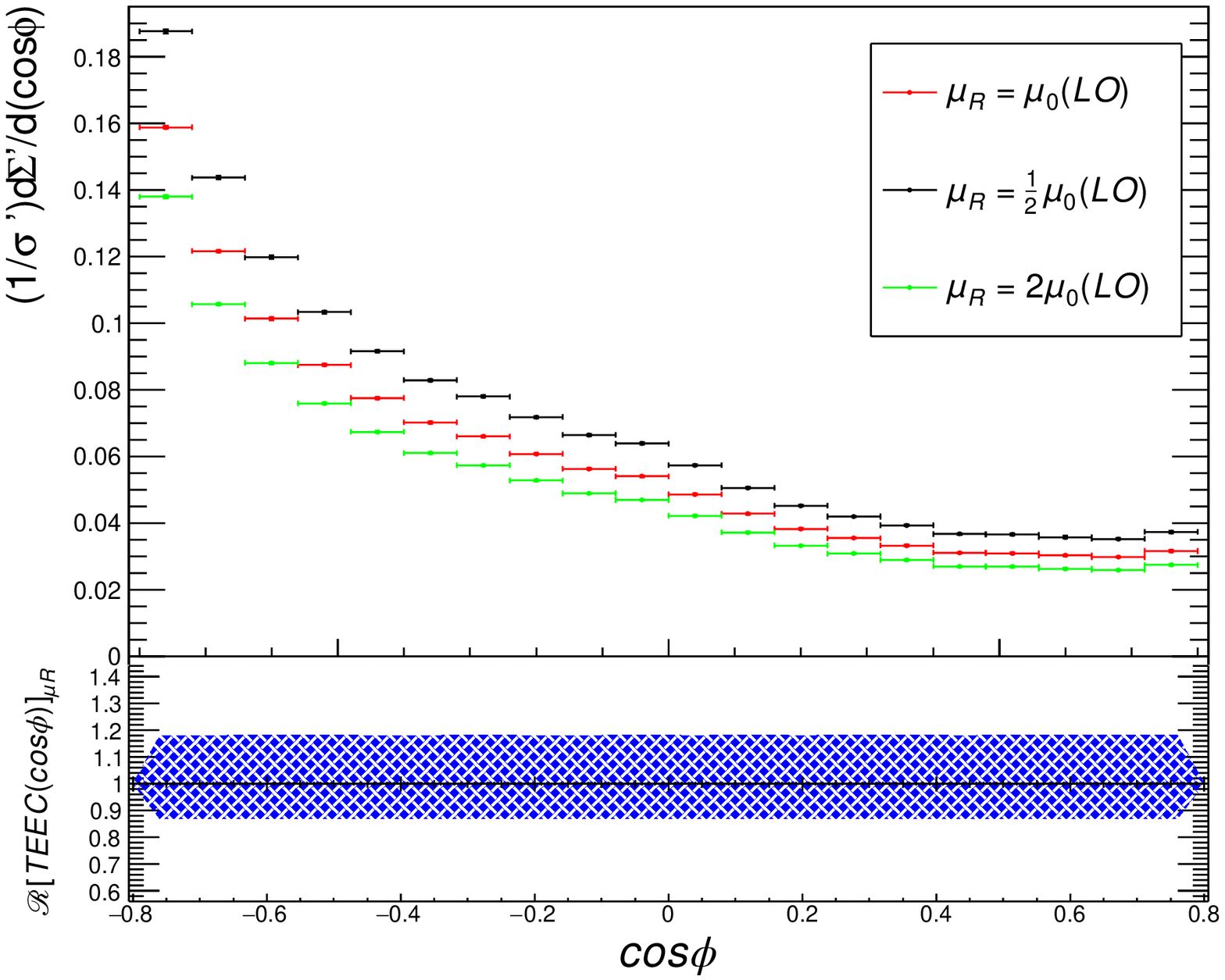}
    \end{minipage}
   \begin{minipage}[t]{0.40\linewidth}
  \centering
  \includegraphics[width=1.0\columnwidth]{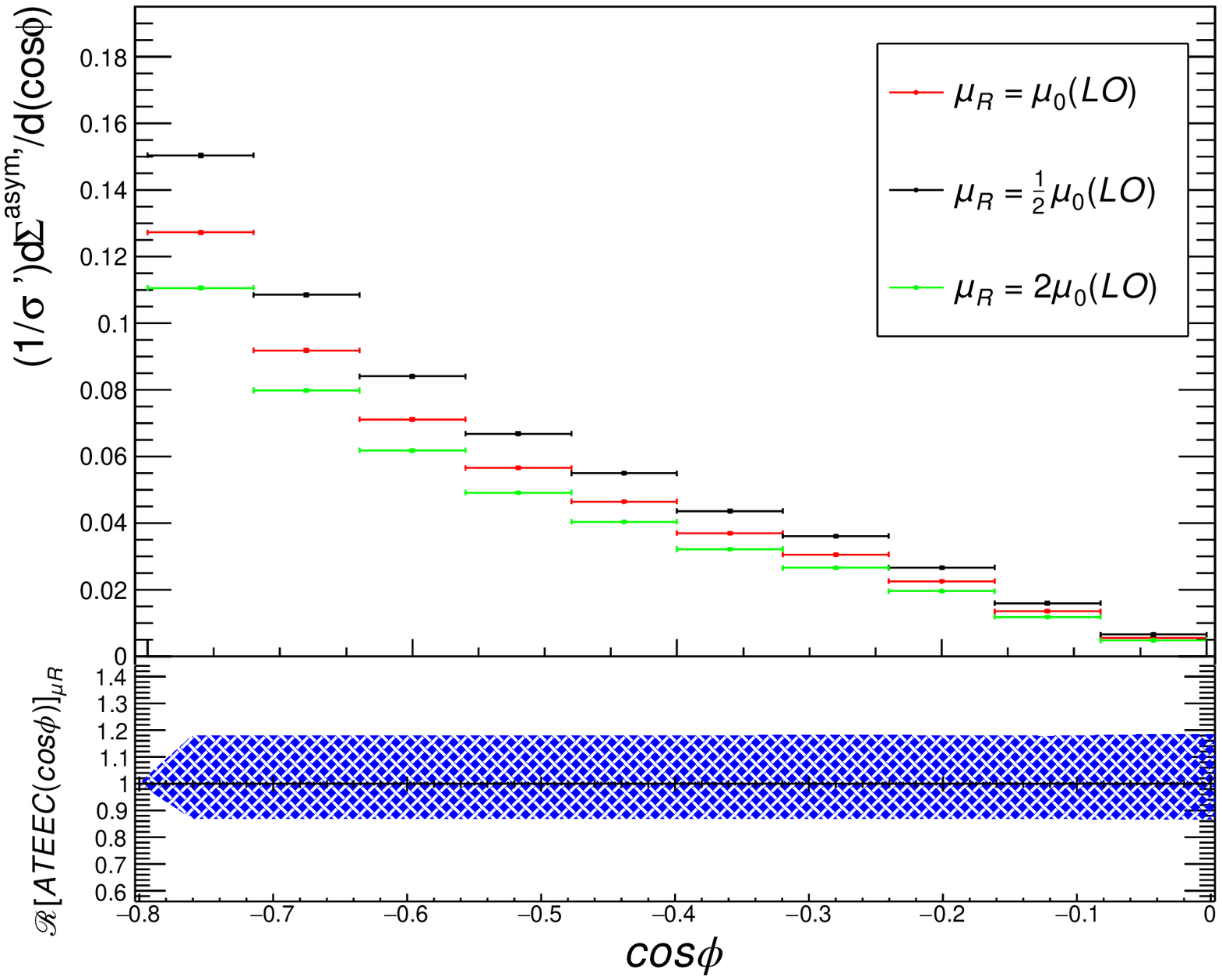}
  \end{minipage}
  \begin{minipage}[t]{0.40\linewidth}
  \centering
  \includegraphics[width=1.0\columnwidth]{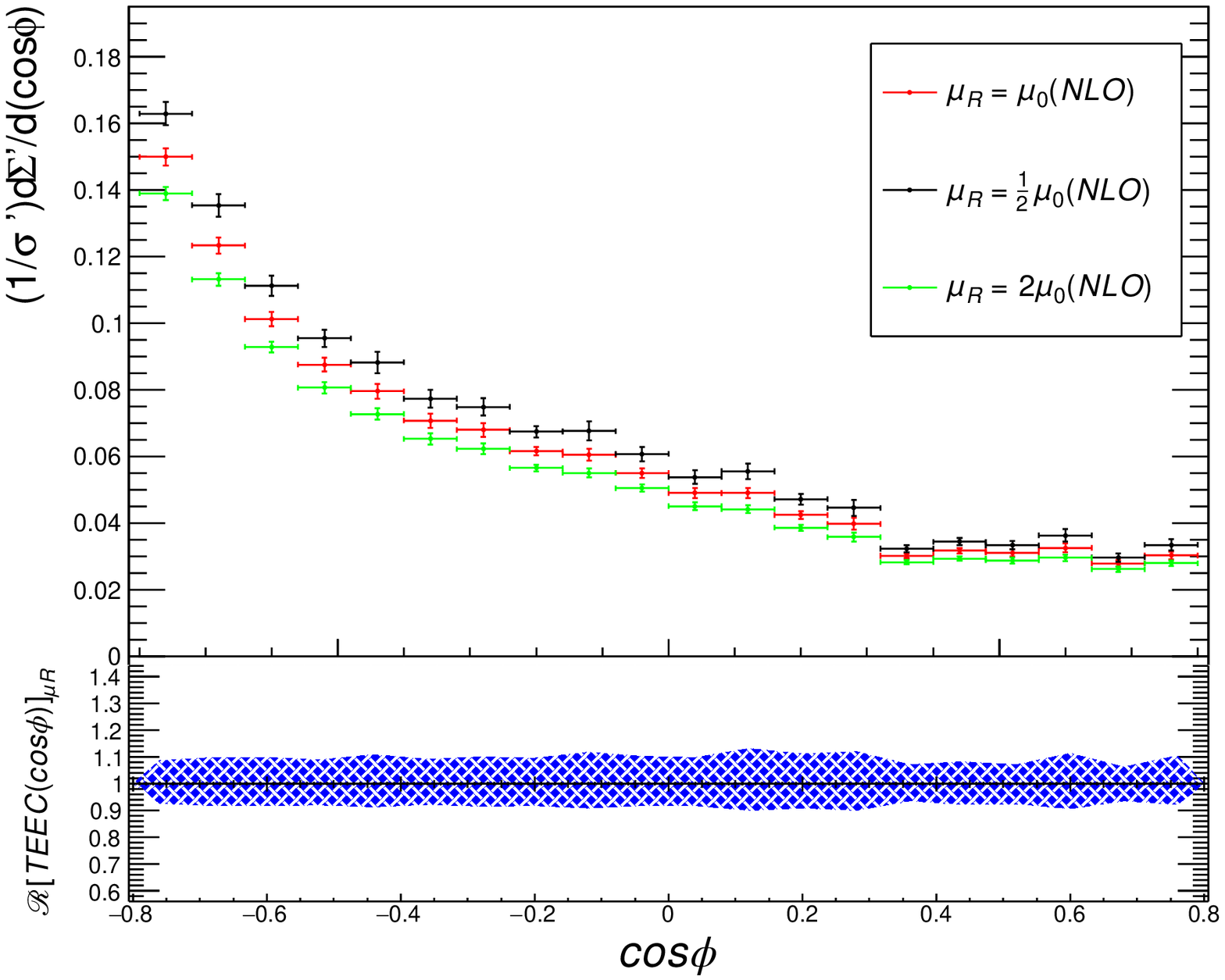}
    \end{minipage}
    \begin{minipage}[t]{0.40\linewidth}
  \centering
  \includegraphics[width=1.0\columnwidth]{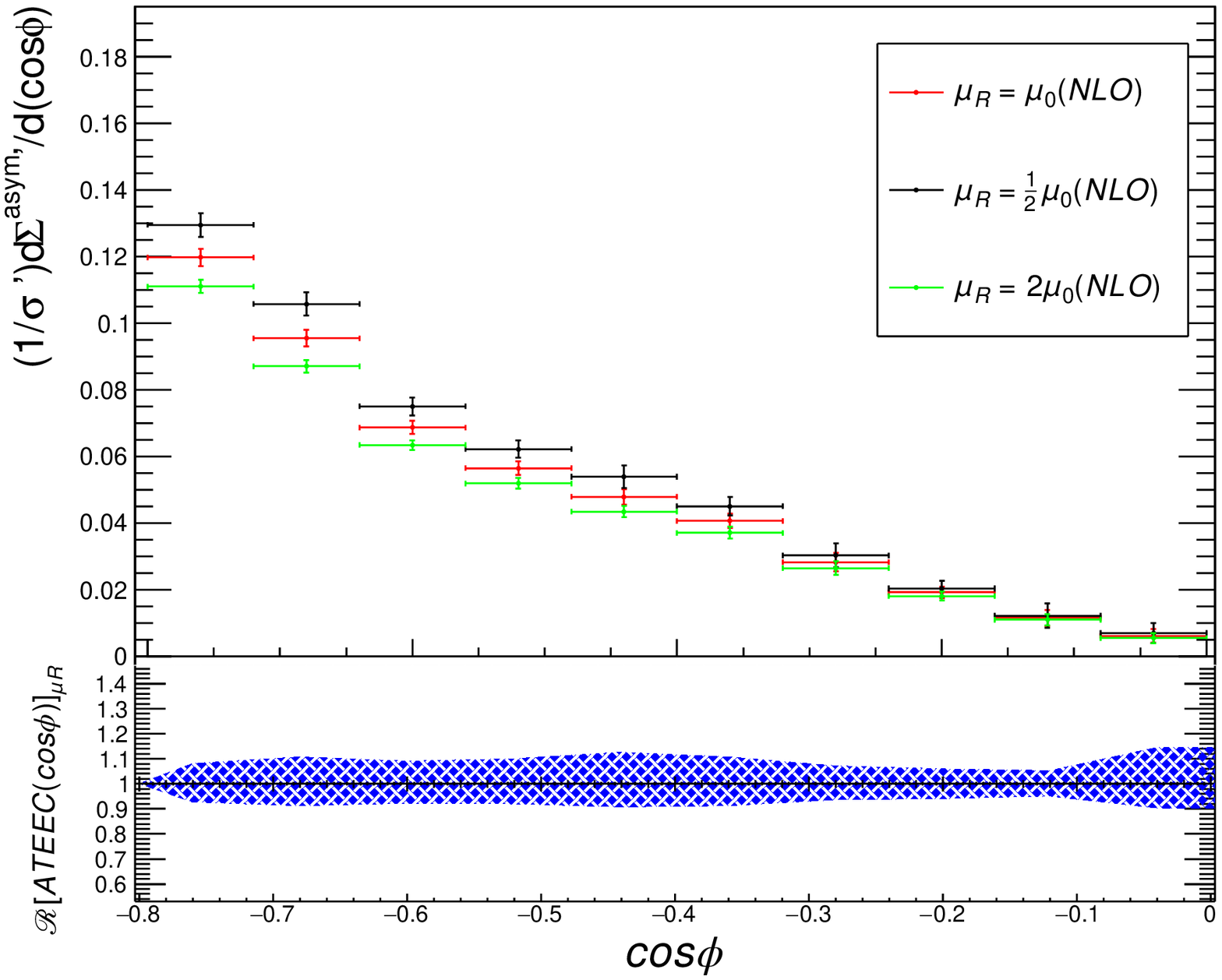}
    \end{minipage}

\caption{Renormalization  scale dependence  of the differential distribution $1/\sigma^\prime d\Sigma^\prime/d (\cos \phi)$  and  its asymmetry
   $ 1/\sigma^\prime d\Sigma^{\prime asym}/d(\cos\phi) $ in the leading order
(upper frames) and the next to leading order (lower frames) varying $\mu_R$ in the range $[0.5,2]\times \mu_0$, where
$\mu_0$ is the nominal scale defined in the text, calculated  with the
MMHT14 PDFs  for the low-$Q^2$: $5.5\,\rm{GeV}^2<Q^2<80\,\rm{GeV}^2$ at HERA. The corresponding $\mu_R$-dependence is also
shown in terms of $ {\mathscr R} [{\rm TEEC} (\cos \phi)]_{\mu_R}$ and $ {\mathscr R} [{\rm ATEEC} (\cos \phi)]_{\mu_R}$,
defined in Eq.~(\ref{eq:dmuR}).
}\label{fig:fig6}
\end{figure}

\begin{figure}[htbp!]
\begin{minipage}[t]{0.40\linewidth}
  \centering
  \includegraphics[width=1.0\columnwidth]{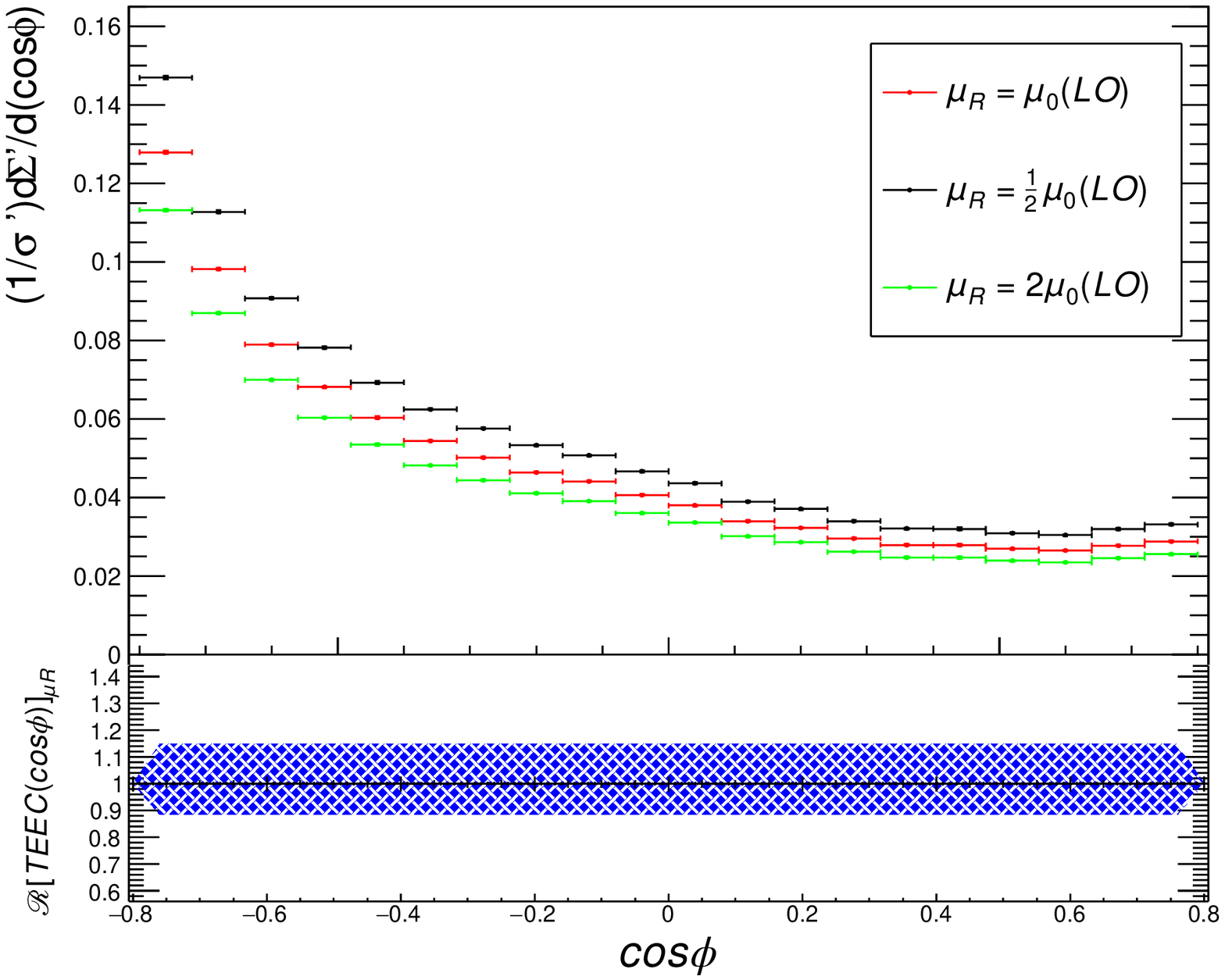}
    \end{minipage}
   \begin{minipage}[t]{0.40\linewidth}
  \centering
  \includegraphics[width=1.0\columnwidth]{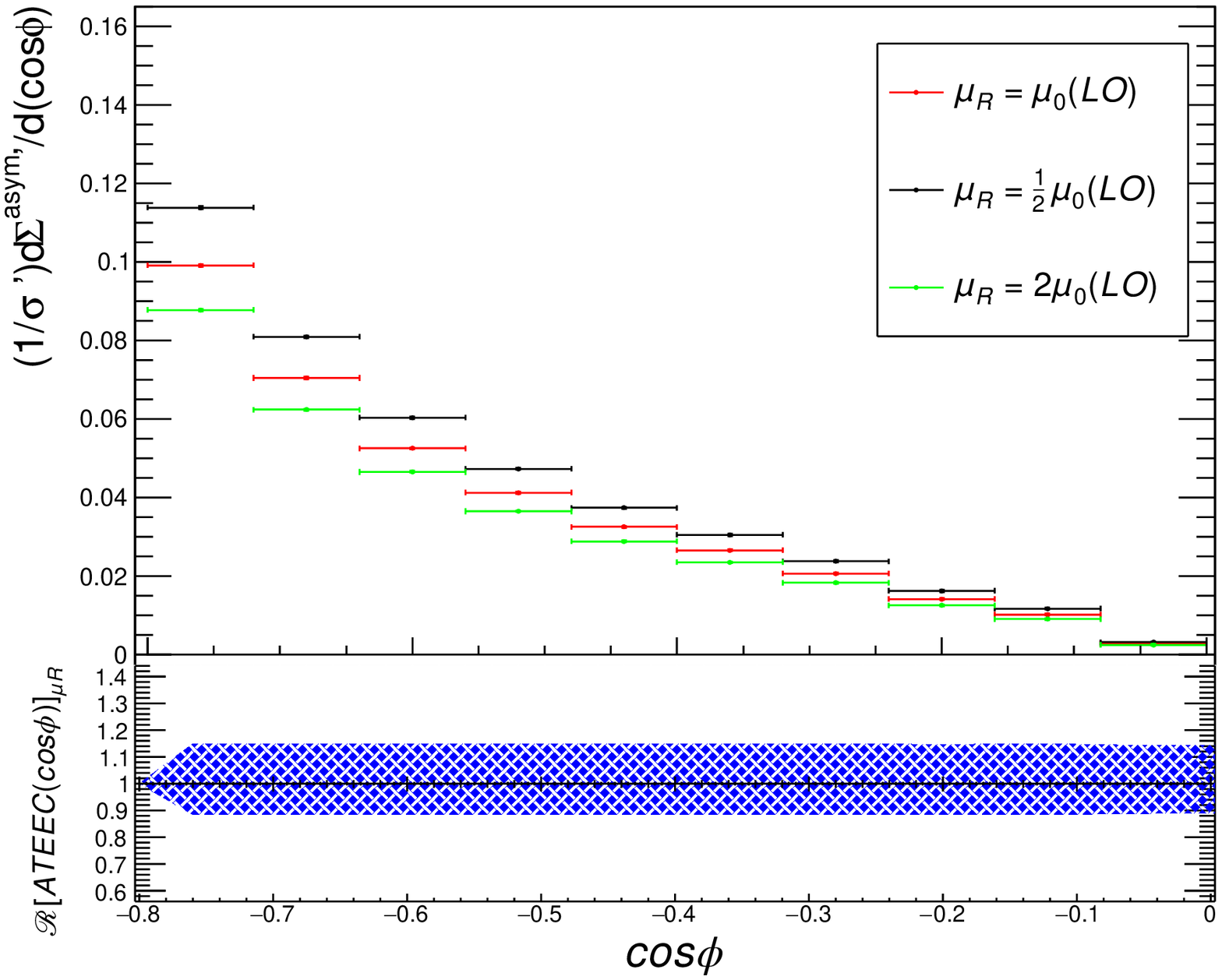}
  \end{minipage}
  \begin{minipage}[t]{0.40\linewidth}
  \centering
  \includegraphics[width=1.0\columnwidth]{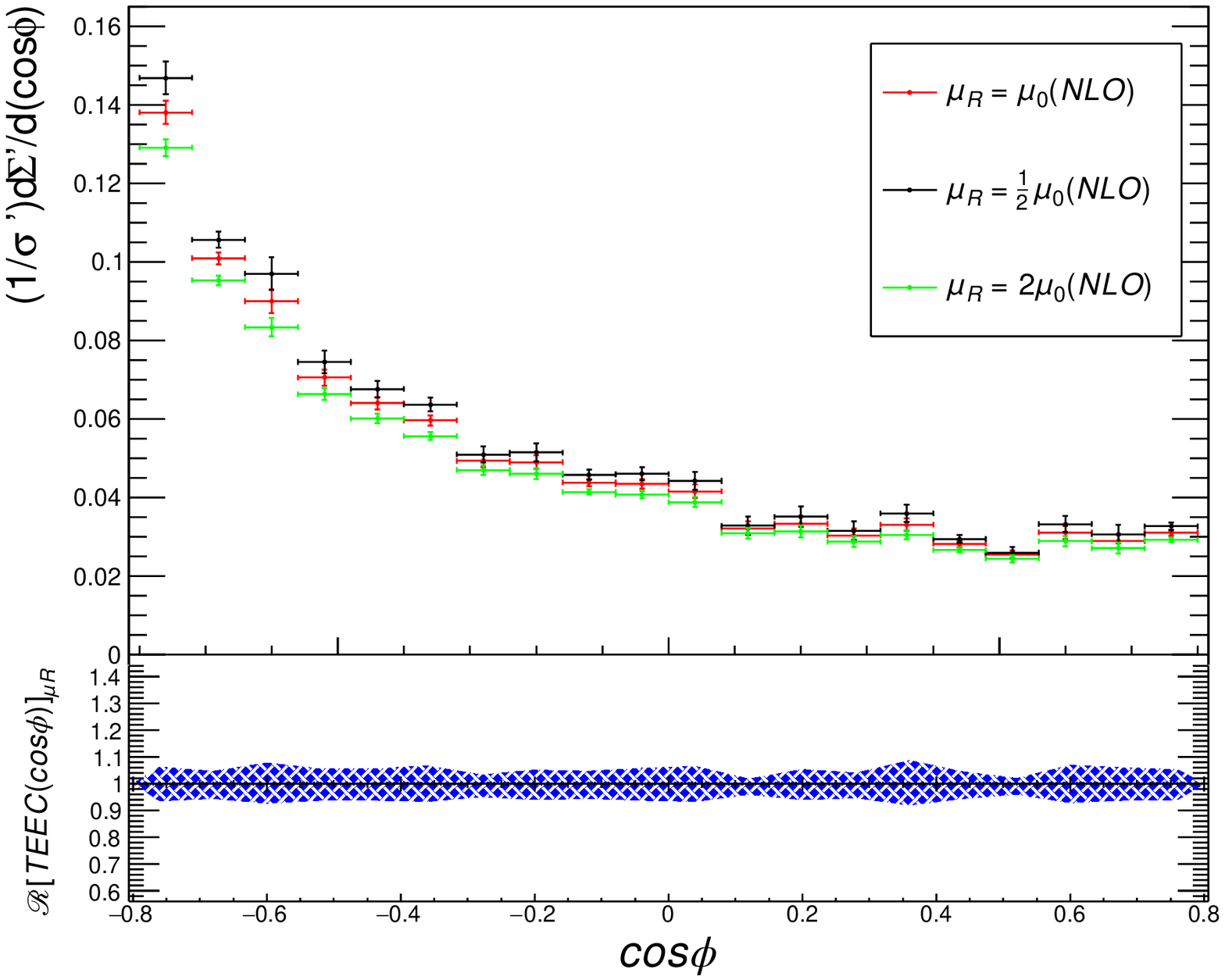}
    \end{minipage}
    \begin{minipage}[t]{0.40\linewidth}
  \centering
  \includegraphics[width=1.0\columnwidth]{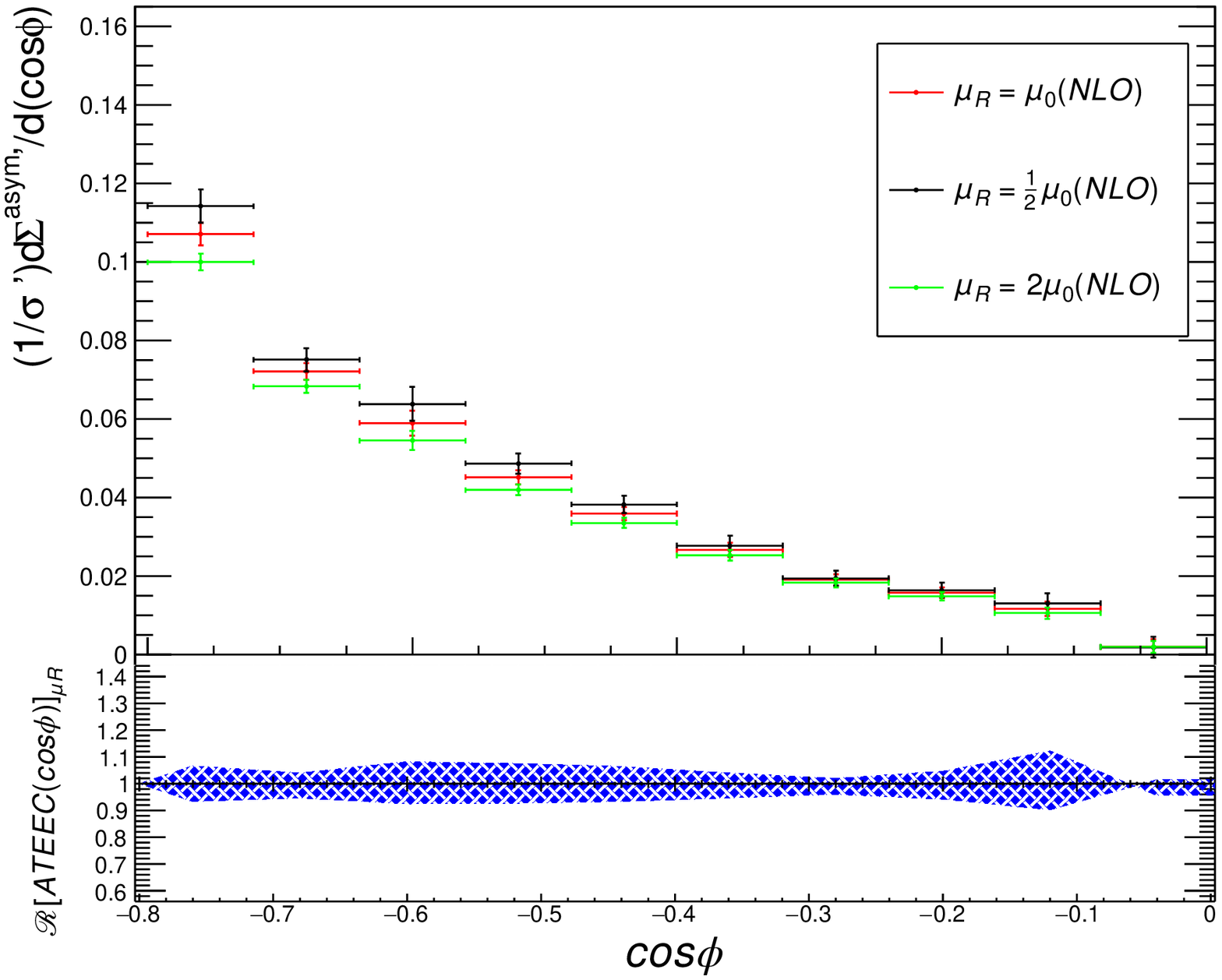}
    \end{minipage}
  \caption{Renormalization  scale dependence  of the differential distribution $1/\sigma^\prime d\Sigma^\prime/d (\cos \phi)$  and  its asymmetry
   $ 1/\sigma^\prime d\Sigma^{\prime asym}/d(\cos\phi) $ in the leading order
(upper frames) and the next to leading order (lower frames) as in Fig.~\ref{fig:fig6}, but for the high-$Q^2$ range:
$150\,\rm{GeV}^2<Q^2<1000\,\rm{GeV}^2$ at HERA.}
\label{fig:fig7}
\end{figure}

The $\mu_R$-dependence
 in the corresponding $Q^2$-ranges are shown in Figs.~\ref{fig:fig6} and~\ref{fig:fig7}, respectively. Here, we fixed $\mu_F=\mu_0$,
 and varied $\mu_R$ in the range $\mu_R=[0.5, 2]\mu_0$. One notices marked improvement in the $\mu_R$-dependence from
 the LO to NLO.
 The $\mu_R$-uncertainty on  TEEC $(\cos \phi)$ and ATEEC $(\cos \phi)$ are  plotted in the lower frames of
 these figures (Figs.~\ref{fig:fig6} and~\ref{fig:fig7})  in terms of the ratios   $ {\mathscr R}$[TEEC$(\cos \phi)]_{\mu_R}$,
 and $ {\mathscr R}$[ATEEC$(\cos \phi)]_{\mu_R}$, defined in Eq.~(\ref{eq:dmuR}).
 \begin{eqnarray}
\label{eq:dmuR}
&&  {\mathscr R}[\text{TEEC}(\cos \phi)]_{\mu_R} \equiv \frac{\text{TEEC}(\cos \phi)_{\mu_R=x \mu_0,\mu_F=\mu_0}} {\text{TEEC}(\cos \phi)_{\mu_R=\mu_F=\mu_0} },\quad x \in [0.5,2],\notag \\
&&
{\mathscr R}[\text{ATEEC}(\cos \phi)]_{\mu_R} \equiv \frac{\text{ATEEC}(\cos \phi)_{\mu_R=x \mu_0,\mu_F=\mu_0} } {\text{ATEEC}(\cos \phi)_{\mu_R=\mu_F=\mu_0}},\quad x \in [0.5,2] .
 \end{eqnarray}
Based on these numerical results, we find that the combined uncertaity due to the PDFs, and the $\mu_F$ and $\mu_R$-scales,
is at about 10\% in the TEEC ($\cos \phi)$, and smaller in ATEEC ($\cos \phi)$.

 Further reduction in the scale uncertainty requires
additional input, which we anticipate from the NNLO improvements as well as from the fits of the
HERA data. This is suggested by the detailed NLO- and NNLO-studies done for the inclusive-jet and dijet data at HERA~\cite{Andreev:2017vxu},
 which can be summarized as follows: The effect of varying $\mu_F$ in the range 10 to 90 GeV on the jet cross sections is small, and this scale can be fixed to a
value within this range without risking a perceptible change elsewhere, which is essentially  in line what we find in our analyis. The effect of varying the scale $\mu_R$ is found more significant in the HERA jet-analysis.
However,  the choice $\mu_R= \sqrt{\langle E_T\rangle^2 +Q^2}$ yields a good fit of the jet data in both the NLO and NNLO accuracy. The reduced $\mu_R$-dependence in the NNLO accuracy leads to a factor 2 improvement in the accuracy of $\alpha_s(M_Z^2)$.  Following~\cite{Andreev:2017vxu}, we shall fix the scale $\mu_R$ to its nominal value in studying the sensitivity of TEEC $(\cos \phi)$ and ATEEC $(\cos \phi)$ on $\alpha_s(M_Z^2)$.

\begin{figure}[htbp!]
  \begin{minipage}[t]{0.40\linewidth}
  \centering
  \includegraphics[width=1.0\columnwidth]{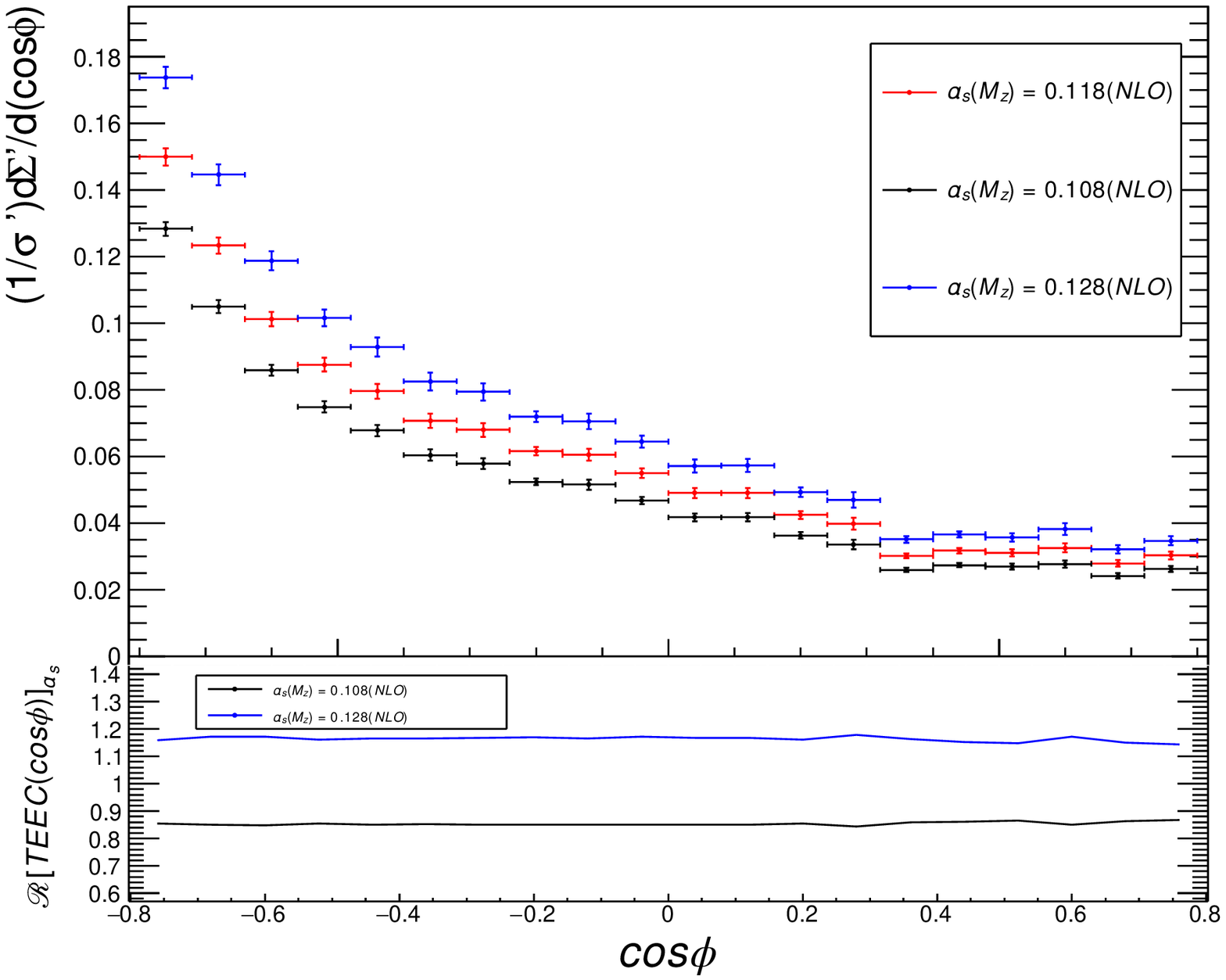}
    \end{minipage}
   \begin{minipage}[t]{0.40\linewidth}
  \centering
  \includegraphics[width=1.0\columnwidth]{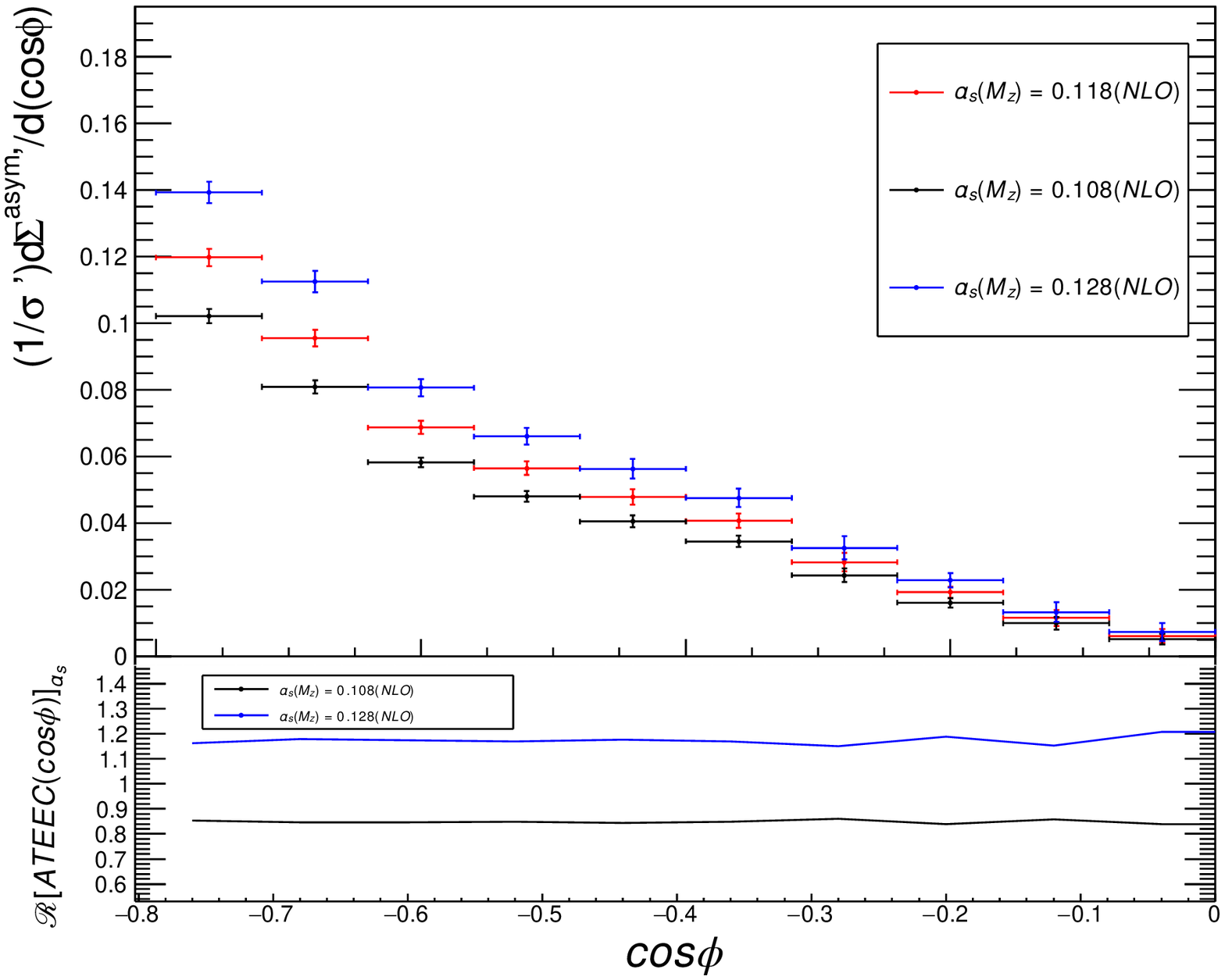}
  \end{minipage}
\caption{
  Upper frames: Dependence of the differential distribution $1/\sigma^\prime d\Sigma^\prime/d (\cos \phi)$  (left) and  its asymmetry
   $ 1/\sigma^\prime d\Sigma^{\prime asym}/d(\cos\phi) $  (right)  on the QCD  coupling constant $\alpha_s(M_Z^2)$
  for three indicated values of $\alpha_s(M_Z^2)= 0.108,0.118,0.128$
  using the PDFs of MMHT14 in the low-$Q^2$ range at HERA  setting  the scales $\mu_F=\mu_R=\mu_0$.
  Lower frames: The ratios $ {\mathscr{R}} [\text{TEEC}(\cos \phi)]_{\alpha_s=0.108}$ and $ {\mathscr{R}} [\text{TEEC}(\cos \phi)]_{\alpha_s=0.128}$
 and  $ {\mathscr{R}} [\text{ATEEC}(\cos \phi)]_{\alpha_s=0.108}$ and $ {\mathscr{R}} [\text{ATEEC}(\cos \phi)]_{\alpha_s=0.128}$,
 defined in Eq.~(\ref{eq:dalphas}).}
  \label{fig:fig8low}
\end{figure}

\begin{figure}[htbp!]
  \begin{minipage}[t]{0.40\linewidth}
  \centering
  \includegraphics[width=1.0\columnwidth]{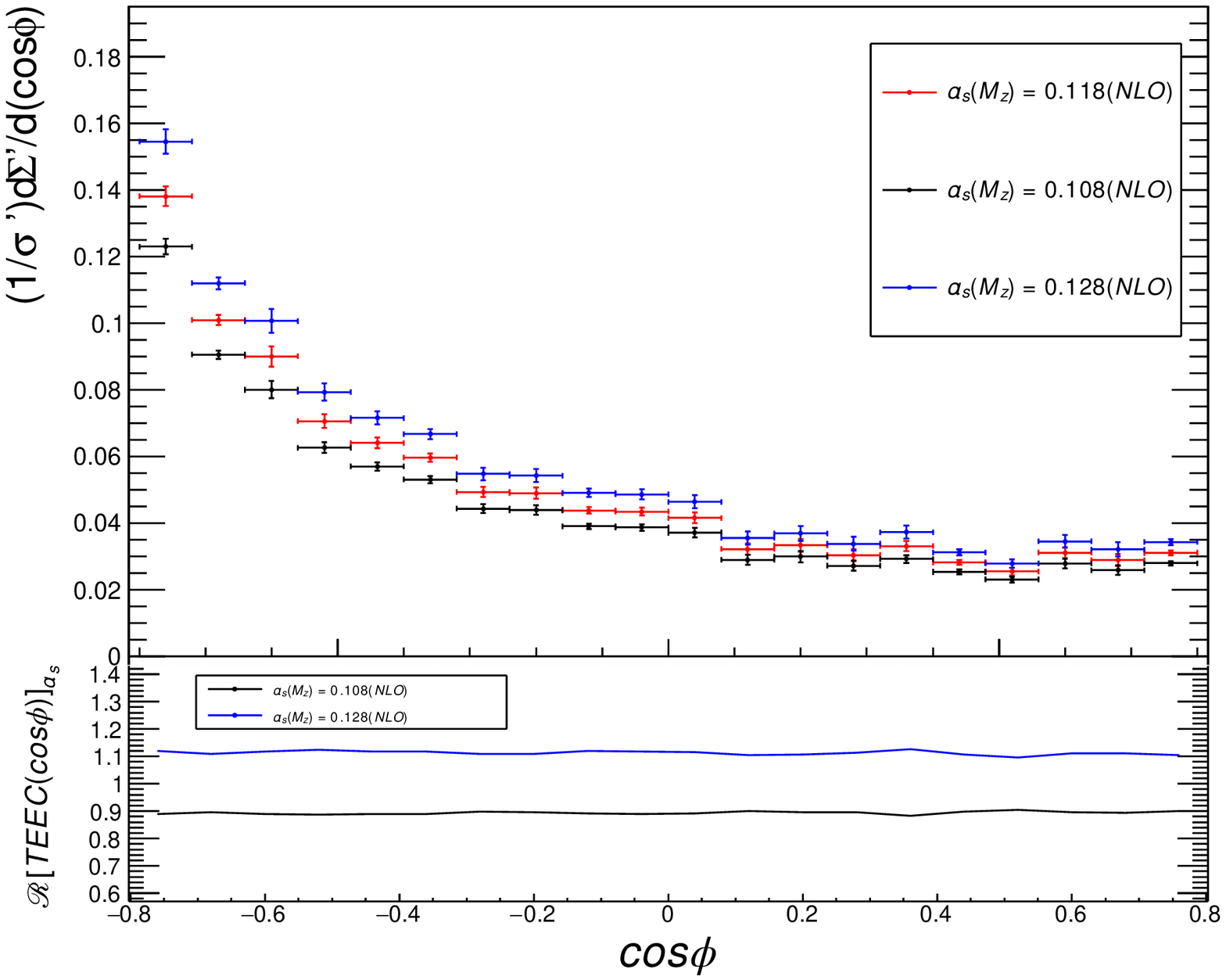}
    \end{minipage}
   \begin{minipage}[t]{0.40\linewidth}
  \centering
  \includegraphics[width=1.0\columnwidth]{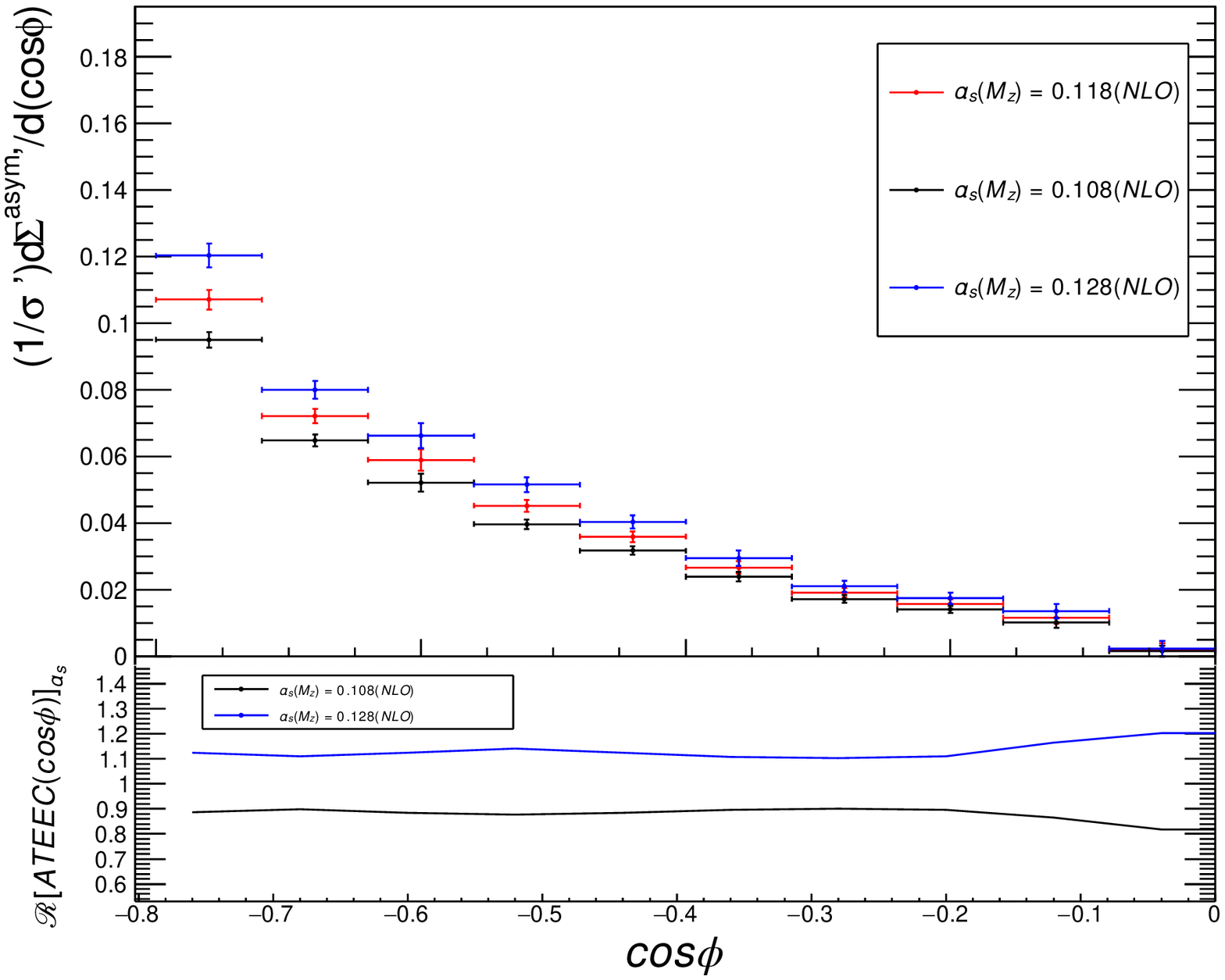}
  \end{minipage}

  \caption{Dependence of the differential distribution $1/\sigma^\prime d\Sigma^\prime/d (\cos \phi)$  (left) and  its asymmetry
   $ 1/\sigma^\prime d\Sigma^{\prime asym}/d(\cos\phi) $  (right)  on the QCD  coupling constant $\alpha_s(M_Z^2)$
  as in Fig.~\ref{fig:fig8low}, but for the  high-$Q^2$ range at HERA.}
  \label{fig:fig9}
\end{figure}

We now discuss the sensitivity of  TEEC ($\cos \phi)$  and ATEEC ($\cos \phi)$   on $\alpha_s(M_Z^2)$.
The results presented are obtained by making the nominal choice of the scales $\mu_F=\mu_R=\mu_0$ and the
 MMHT14 PDFs.  Results for three representative values
$\alpha_s(M_Z^2)=0.108,\,0.118,\,0.128$ are shown, which bracket most other determinations of this quantity, with $\alpha_s(M_Z^2)=0.118$ being
the central value\footnote{The current PDG world average is $\alpha_s(M_Z^2)=0.1179 \pm 0.0010$.}
 quoted by the Particle Data Group~\cite{Zyla:2020zbs}. They are shown in Fig.~\ref{fig:fig8low} (low$-Q^2$ range) and
Fig.~\ref{fig:fig9} (high$-Q^2$ range) at the NLO accuracy.  To quantify the
$\alpha_s(M_Z^2)$-sensitivity, we define the following ratios:
\begin{eqnarray}
\label{eq:dalphas}
&&   {\mathscr R} [\text{TEEC}(\cos \phi)]_{\alpha_s} \equiv  \frac{\text{TEEC}(\cos \phi)_{\alpha_s(M_Z^2)}} {\text{TEEC}(\cos \phi)_{\alpha_s(M_Z^2)=0.118} } \notag \\
&&
 {\mathscr R} [\text{ATEEC}(\cos \phi)]_{\alpha_s} \equiv \frac{\text{ATEEC}(\cos \phi)_{\alpha_s(M_Z^2)}} {\text{ATEEC}(\cos \phi)_{\alpha_s(M_Z^2)=0.118}}.
 \end{eqnarray}

They are shown in the bottom frames in Fig.~\ref{fig:fig8low} (low$-Q^2$ range) and  Fig.~\ref{fig:fig9} (high$-Q^2$ range) in
terms of the ratios ${\mathscr R} [\text{TEEC}(\cos \phi)]_{\alpha_s=0.128}$ and ${\mathscr R} [\text{TEEC}(\cos \phi)]_{\alpha_s=0.108}$.
 We note that both $ {\mathscr R}[\text{TEEC}(\cos \phi)]_{\alpha_s}$ and  the corresponding ratio for the asymmetry $ {\mathscr R}[\text{ATEEC}(\cos \phi)]_{\alpha_s}$
show a marked sensitivity  on $\alpha_s(M_Z^2)$.
Hence, these shape functions at HERA offer competitive avenues to determine
$\alpha_s(M_Z^2)$, and we urge our experimental colleagues to undertake a detailed data analysis of these variables at HERA.

\begin{figure}[htbp!]
  \begin{minipage}[t]{0.40\linewidth}
  \centering
  \includegraphics[width=1.0\columnwidth]{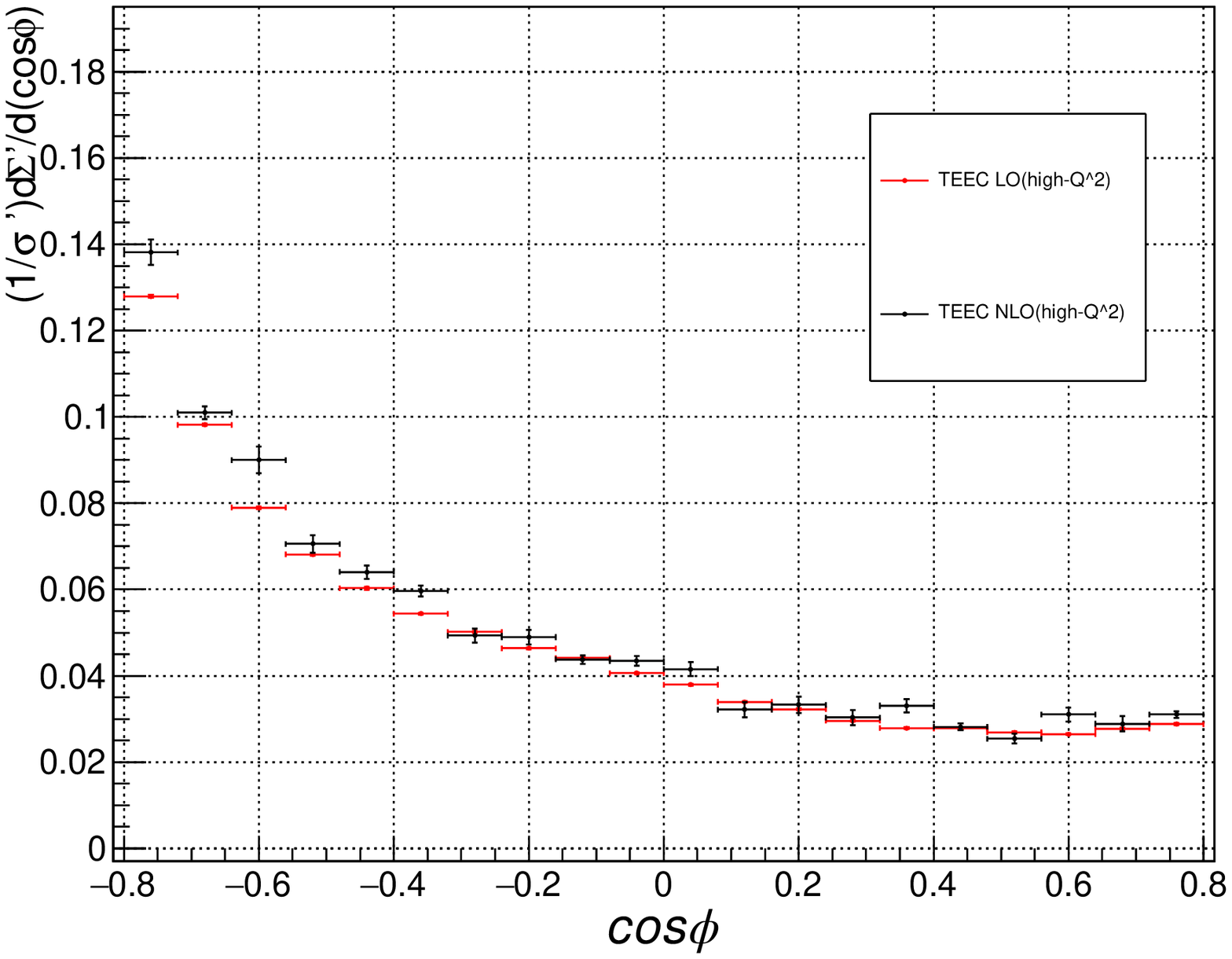}
    \end{minipage}
   \begin{minipage}[t]{0.40\linewidth}
  \centering
  \includegraphics[width=1.0\columnwidth]{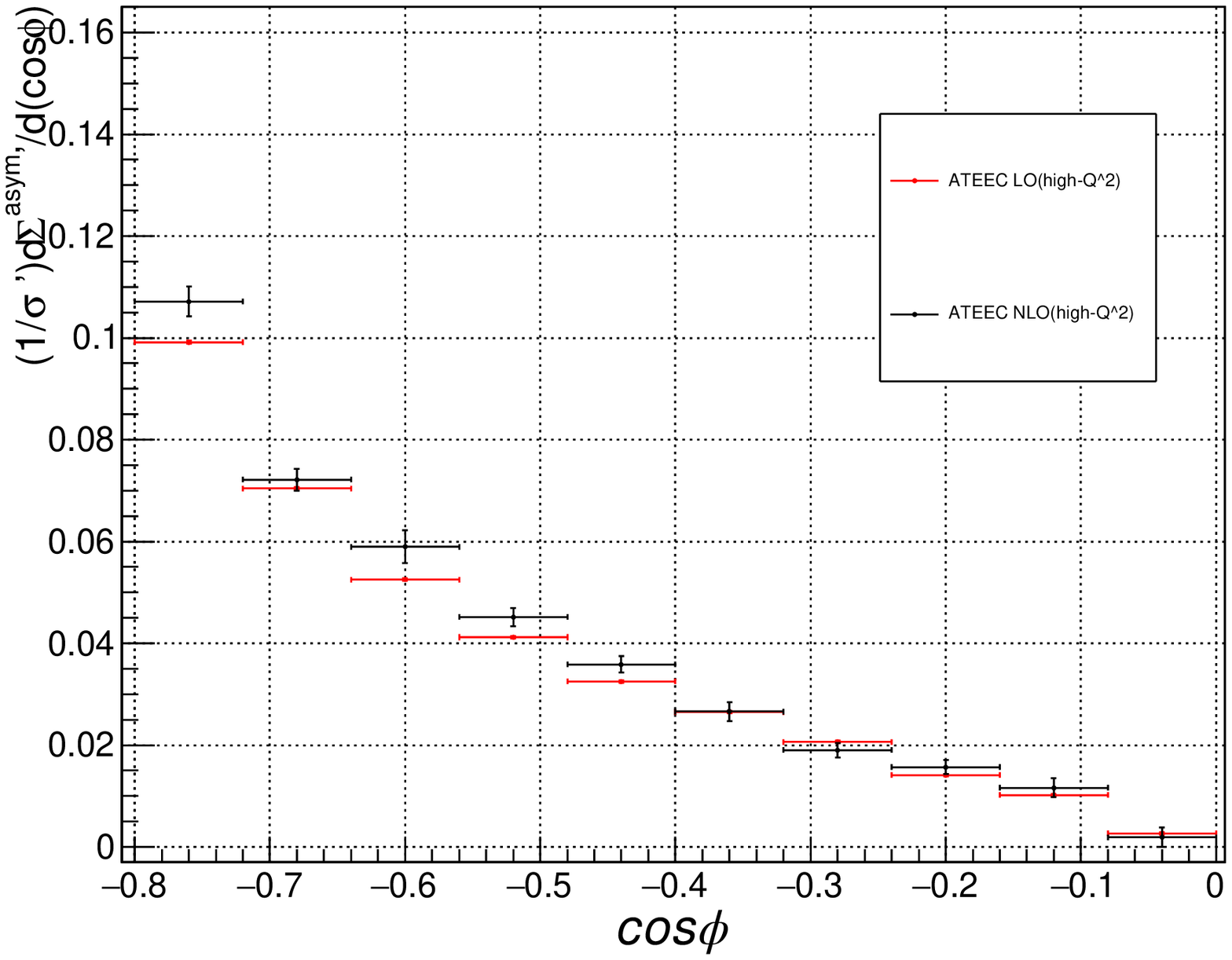}
  \end{minipage}
  \begin{minipage}[t]{0.40\linewidth}
  \centering
  \includegraphics[width=1.0\columnwidth]{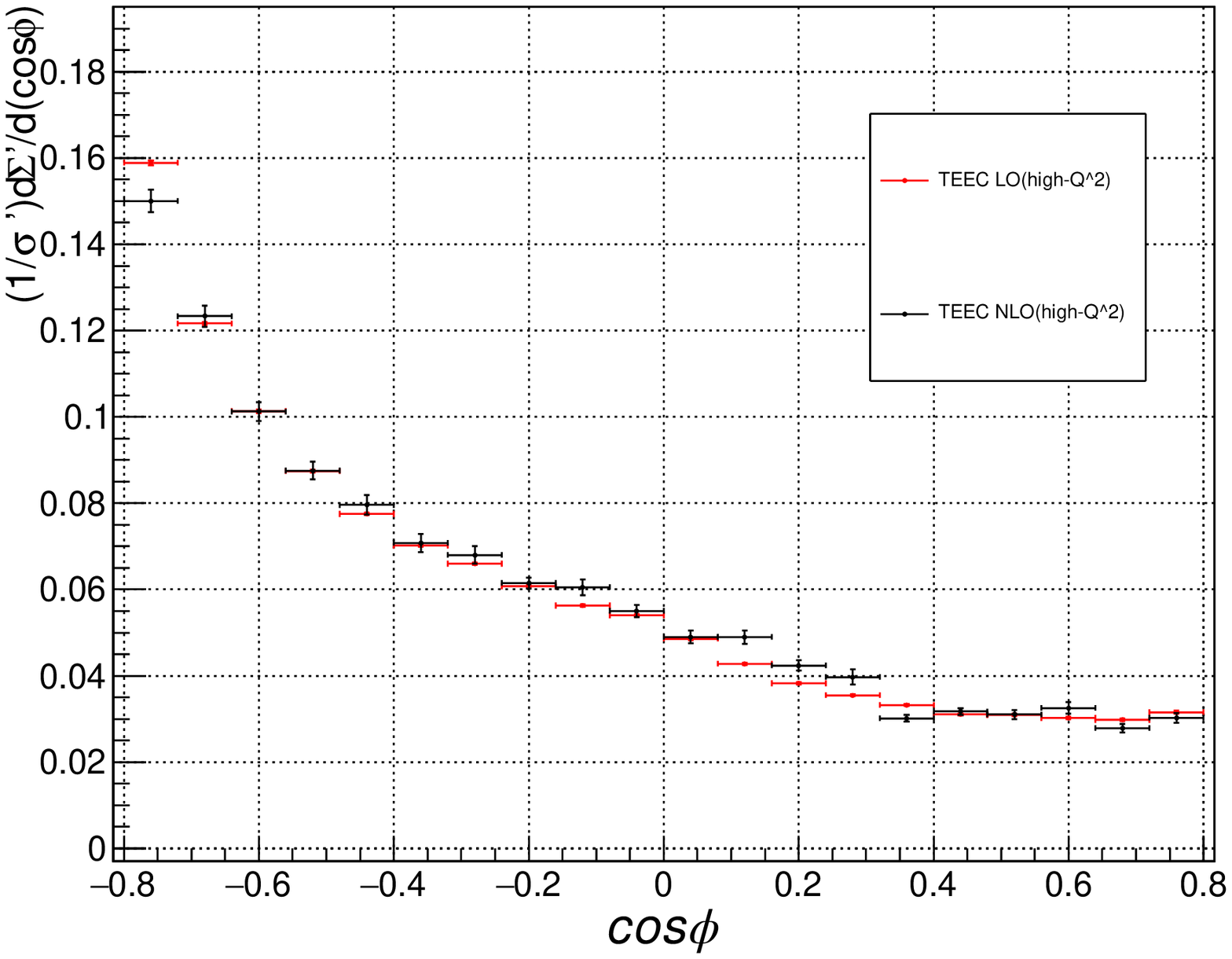}
    \end{minipage}
    \begin{minipage}[t]{0.40\linewidth}
  \centering
  \includegraphics[width=1.0\columnwidth]{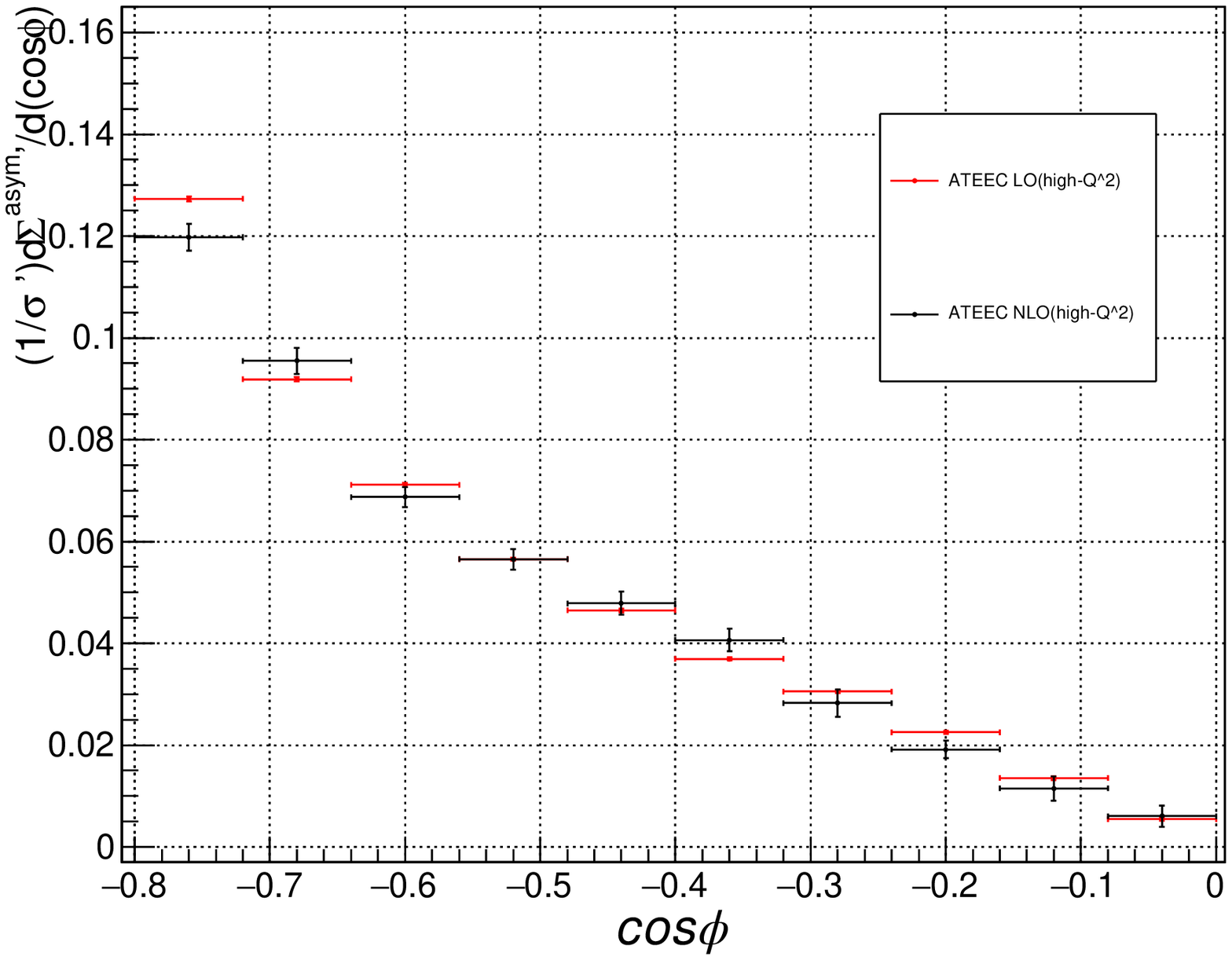}
    \end{minipage}
\caption{A comparison of the LO and the NLO differential distribution $1/\sigma^\prime d\Sigma^\prime/d (\cos \phi)$  (left) and  its asymmetry
   $ 1/\sigma^\prime d\Sigma^{\prime asym}/d(\cos\phi) $  (right)
at HERA ($\sqrt{s}=314$ GeV) in the high-$Q^2$ range
(upper frames) and low-$Q^2$ range (lower frames), with $\mu_F=\mu_R=\mu_0=\sqrt{\langle E_T\rangle^2+Q^2}$ and $\alpha_s(M_Z^2)=0.118$. }\label{fig:fig1}
\end{figure}

 A comparison of the LO and the NLO TEEC$(\cos \phi)$  and its asymmetry  ATEEC $(\cos \phi)$ at HERA ($\sqrt{s}=314$ GeV) in the high-$Q^2$ range and
 the low-$Q^2$ range are shown in Fig.~\ref{fig:fig1}. These results are obtained for the choice $\mu_F=\mu_R=\mu_0=\sqrt{\langle E_T\rangle^2+Q^2}$, $\alpha_s(M_Z^2)=0.118$, and MMHT14 set of PDFs. They show that theses correlations are remarkably stable against
 NLO corrections. We conjecture that NNLO corrections are, likewise, small. This remains to be shown and we hope that our work will
 stimulate working them out.

\section{Summary}

In this paper,
we have  studied for the first time, the transverse energy-energy correlations TEEC $(\cos \phi)$ and its asymmetry ATEEC $(\cos \phi)$ in
deep inelastic scattering at the electron-proton collider HERA at the center of mass energy $\sqrt{s}=314$ GeV, where  $\phi$ is the angle in the Breit frame between two jets defined using a transverse-momentum $(k_T)$ jet algorithm. We use  NLOJET++ to calculate  these functions in the LO and the  NLO approximations in QCD for two ranges in the momentum transfer squared $Q^2$.  In the LO, these results are checked using the package \text{MadGraph5\_aMC@NLO}~\cite{Alwall:2014hca}  with the MMHT14  PDF set.
 We show the sensitivity of these functions on the PDFs, factorization $(\mu_F)$ and renormalization $(\mu_R)$ scales, and  on $\alpha_s(M_Z^2)$.  With the various cuts in the event generation matched with  the ones in the measurements by the H1 collaboration at HERA, these studies are potentially useful  in the analysis of the HERA data, including the determination of $\alpha_s(M_Z^2)$ from the shape variables.

 An NNLO calculation  for these shape variables is still lacking. This has the consequence that significant
renormalization-scale dependence which enters in the partonic cross sections remains. At the present theoretical accuracy followed in this
paper, this may compromise the precision on $\alpha_s(M_Z^2)$.  Theoretical precision can be improved by including the
NNLO contribution, as shown for the dijet and inclusive jet cross sections in in
 DIS~\cite{Klasen:2013cba,Biekotter:2015nra,Currie:2016ytq,Currie:2017tpe}.
However, the scale uncertainty  could also be reduced by analysing the HERA data for the shape variables by
narrowing the allowed range of $\mu_R$ for which one gets a good quality fit. This is the case in the analysis of the inclusive-jet and dijet HERA
data, in which the choice $\mu_R=\sqrt{\langle E_T\rangle^2 +Q^2}$ accounts well for the H1 measurements, also in the NLO accuracy~\cite{Andreev:2017vxu}. For this choice  of the $\mu_R$ scale, we have shown that the event shape TEEC $( \cos \phi )$ and its asymmetry are very sensitive to the value of $\alpha_s(M_Z^2)$. We hope that our case-study for the TEEC and ATEEC at HERA, carried out at the NLO accuracy,  will help fous on the analysis of the data on these shape vaiables with improved theoretical accuracy.

\section*{Acknowledgements}
We thank Fernando Barreiro,  Xiao-Hui Liu and  Stefan Schmitt for valuable discussions on the HERA measurements of jets.
This work is supported in part by the Natural Science Foundation of China under grant No.   11735010, 11911530088, U2032102,  by the Natural Science Foundation of Shanghai under grant No. 15DZ2272100, and by MOE Key Lab for Particle Physics, Astrophysics and Cosmology. GL is supported under U.S. Department of Energy contract DE-SC0011095.

\section*{Appendix-A}
\begin{figure}[htbp!]
   \begin{minipage}[t]{0.45\linewidth}
  \centering
  \includegraphics[width=1.0\columnwidth]{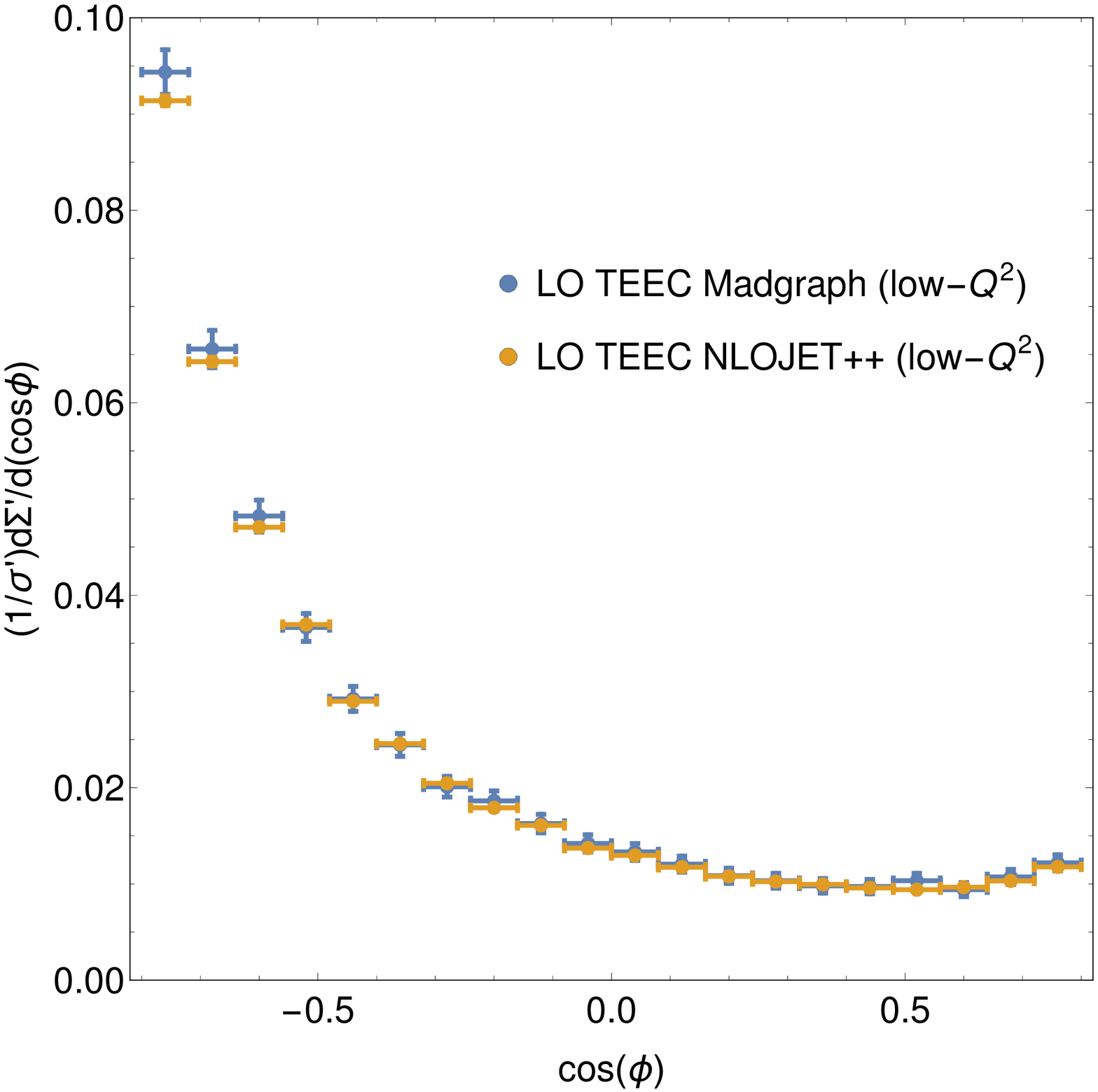}
  \end{minipage}
   \begin{minipage}[t]{0.45\linewidth}
  \centering
  \includegraphics[width=1.0\columnwidth]{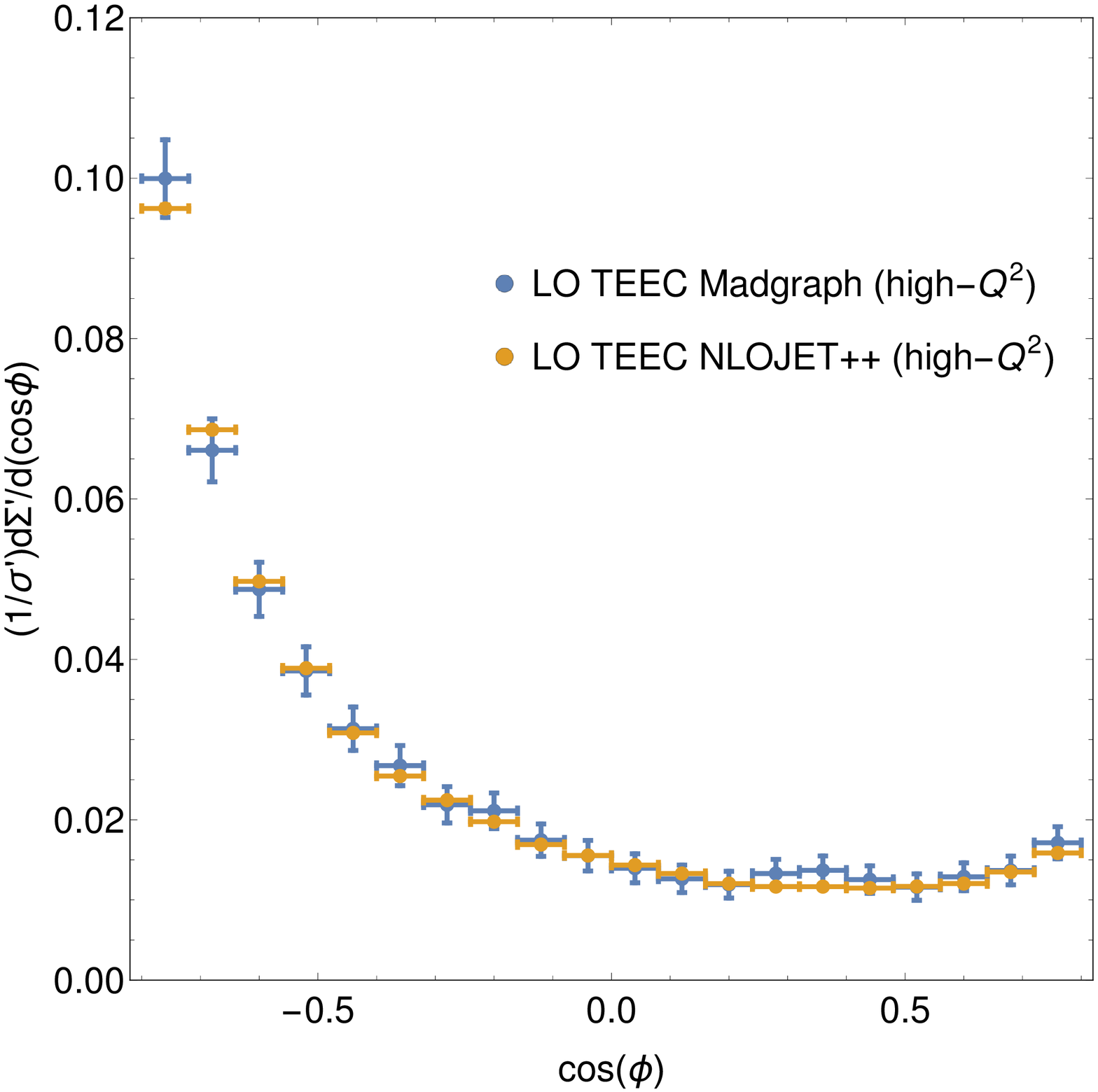}
  \end{minipage}
\caption{Comparison of the LO differential distribution $1/\sigma^\prime d\Sigma^\prime/d (\cos \phi)$
obtained using Madgraph and NLOJET++ in the low-$Q^2$ range (left frame)
and the high-$Q^2$ range (right frame). }
\label{fig:EEC-MG5}
\end{figure}

As a cross check on our calculations, we have also used the program  Madgraph to calculate the leading order TEEC and ATEEC functions.
To compare with the results obtained using \textsc{NLOJET++}, parton-level events are generated in \text{MadGraph5\_aMC@NLO}~\cite{Alwall:2014hca}  with the MMHT14  PDF set. To that end, the following basic cuts in the lab frame are imposed at the generator level in Madgraph:
\begin{align}
\label{eq:basic cuts}
p_{T,j}^{\text{lab}}>2~\text{GeV},\quad |\eta_{j,e}^{\text{lab}}|<5,\quad \Delta R_{jj}^{\text{lab}}>0.1.
\end{align}
In the above,  $j$ denotes light-flavor quarks, and the angular distance in the $\eta-\phi$ plane is defined as $\Delta R_{ij}\equiv\sqrt{(\eta_i-\eta_j)^2+(\phi_i-\phi_j)^2}$ with $\eta_i$ and $\phi_i$ being the pseudo-rapidity and azimuthal angle of particle $i$, respectively.
The momenta of the generated events are defined in the lab frame. After the appropriate Lorentz transformation, the TEEC and ATEEC distributions in the Breit frame can be constructed, and the events are selected in the low and  high $Q^2$ ranges.
Given the available  choices of the factorization and renormalization scales in Madgraph, we  set the scales $\mu_F=\mu_R=E_T$ with $E_T$ being the scalar sum of transverse energies of all jets in both Madgraph and NLOJET++. The transverse energy of each jet in the Breit frame is limited in the range $[4.5~\text{GeV},50~\text{GeV}]$ ~\cite{Andreev:2016tgi} to reduce the impact of the basic cuts in Eq.~\eqref{eq:basic cuts}, apart from the cuts on the transverse energies for dijet and trijet events in Eq.~\eqref{eq:etcuts}. In Fig.~\ref{fig:EEC-MG5}, a comparison of the LO TEEC $(\cos\phi)$ distributions obtained using NLOJET++ and Madgraph is shown, using
$\alpha_s(M_Z^2) =0.118$. The distributions obtained from the two packages agree well in both the low-$Q^2$ and high-$Q^2$ ranges. In Fig.~\ref{fig:EEC-MG5} the error bars indicate the statistical errors, which are negligible for the distributions obtained from NLOJET++.

\end{document}